\documentclass[twocolumn]{aastex631}
\usepackage{amsmath}
\usepackage{cuted}

\usepackage{enumitem}
\begin{document}

\title{Dynamical Origin of the Vertical Metallicity Gradient of the Milky Way Bulge}

\author[0000-0001-8962-663X]{Bin-Hui Chen}
\affiliation{Tsung-Dao Lee Institute, Shanghai Jiao Tong University, Shanghai 200240, People’s Republic of China}
\affiliation{Department of Astronomy, School of Physics and Astronomy, Shanghai Jiao Tong University, 800 Dongchuan Road, Shanghai 200240, People’s Republic of China}
\affiliation{Key Laboratory for Particle Astrophysics and Cosmology (MOE)/Shanghai Key Laboratory for Particle Physics and Cosmology, Shanghai 200240, People’s Republic of China}

\author[0000-0001-5604-1643]{Juntai Shen}
\correspondingauthor{Juntai Shen}
\email{jtshen@sjtu.edu.cn}
\affiliation{Department of Astronomy, School of Physics and Astronomy, Shanghai Jiao Tong University, 800 Dongchuan Road, Shanghai 200240, People’s Republic of China}
\affiliation{Key Laboratory for Particle Astrophysics and Cosmology (MOE)/Shanghai Key Laboratory for Particle Physics and Cosmology, Shanghai 200240, People’s Republic of China}

\author{Zhong Liu}
\affiliation{Shanghai Astronomical Observatory, Chinese Academy of Sciences, 80 Nandan Road, Shanghai 200030, P.R. China}



\begin{abstract}

A vertical metallicity gradient in the Milky Way bulge is well-established. Yet, its origin has not been fully understood under the Galactic secular evolution scenario. We construct single-disk and triple-disk $N$-body models with an initial radial metallicity gradient for each disk. These models generate a vertical metallicity gradient through a ``two-step heating" mechanism: first the outer, metal-poor particles move inward via the bar instability and subsequently undergo more significant vertical heating during the buckling instability, so they end up at greater vertical height. The ``two-step heating" mechanism nearly linearly transforms the radial metallicity gradients in precursor disks into vertical metallicity gradients. Comparing the models with a triple-disk model tagged with radially independent Gaussian metallicity, we find that, despite certain limitations, the ``two-step heating" mechanism is still important in shaping the Galactic vertical metallicity gradient. If the bar and buckling instabilities contributed to the formation of boxy/peanut-shaped bulges, then the ``two-step heating" mechanism is inevitable in the secular evolution of a boxy/peanut-shaped bulge.

\end{abstract}




\def\sAlphaR0{-0.33} 
\def\sMH0{0.55} 
\def\sMinorVMG{-0.044} 
\def\sFinalRMG{-0.2} 
\def\stDuration{6.0} 
\def\sDiskPartNum{$5\,000\,000$}
\def\sHaloPartNum{$5\,000\,000$}

\def\ttDuration{5.556} 
\def\tThinPartNum{$2\,100\,000$}
\def\tInterPartNum{$1\,300\,000$}
\def\tThickPartNum{$950\,000$}
\def\tHaloPartNum{$4\,000\,000$}
\def\tThinMHmean{0.3}
\def\tInterMHmean{-0.2}
\def\tThickMHmean{-0.6}
\def\thinAlphaR0{-0.25} 
\def\thinMH0{0.60} 
\def\interAlphaR0{-0.28}
\def\interMH0{0.45}
\def\thickAlphaR0{-0.33} 
\def\thickMH0{0.40} 
\def\tMinorVMG{-0.035} 

\section{Introduction} \label{sec: intro}

Disk galaxies form during hierarchical merging in the early universe. After the galaxy virialized at the galactic scale, its disk-like morphology is set up by the spin-up of the whole galaxy \citep{mo_etal_1998}, through either strong gas accretion \citep{tacche_etal_2016}, and/or clump scenario \citep{noguch_1999, immeli_etal_2004}.

The Milky Way (MW) galaxy is probably a typical disk galaxy. Recent studies of the Galactic Bulge have greatly improved our understanding of the formation and evolution of the MW. The observed photometric and kinematic properties in the Galactic center reveal the disk origin of the MW, including the asymmetric parallelogram shape \citep{bli_spe_1991, weilan_etal_1994, binney_etal_1997}, cylindrical rotation \citep{rich_etal_2007, howard_etal_2009, kunder_etal_2012, ness_etal_2013_a, wylie_etal_2021}, and ``X"-shaped structure \citep{mcw_zoc_2010, nataf_etal_2010, weg_ger_2013, nes_lan_2016}. Disk galaxy simulations in which a boxy/peanut-shaped bulge forms as a result of the internal evolution in the disk can reproduce these results well \citep{shen_etal_2010, li_she_2012}. This well-established bar-made bulge formation scenario suggests that the bulk of the Galactic Bulge originates from a precursor disk instead of a prominent classical bulge \citep{dimatt_2016, bla_ger_2016, she_zhe_2020}.

The complex chemical properties of the Galactic Bulge observed in many spectroscopic surveys (e.g., Gaia-ESO, GIBS, VVV, and ARGOS) still pose challenges to this scenario. The vertical metallicity gradient (VMG) of the Galactic Bulge \citep{zoccal_etal_2008, johnso_etal_2011, gonzal_etal_2013, ness_etal_2013_b, rojas_etal_2014} was once regarded as solid evidence against a boxy/peanut-shaped/``X"-shaped (BPX) bulge. It was once widely believed that a secularly-evolved bar/bulge model tends to mix stars vertically during the buckling process, so such a model could not retain a VMG comparable to observations of the Galaxy \citep{zoccal_etal_2008, babusi_etal_2010}. This belief seems to be confirmed by some secular evolution models with a pure disk, where strong flattening of the pre-existing VMG is reported \citep{friedl_1998, bek_tsu_etal_2011}.

However, \cite{mar_ger_2013} successfully reproduced a VMG in their secularly evolved disk galaxy model. They assume the precursor disk has no pre-existing VMG but only an initial radial metallicity gradient (RMG)\footnote{It is assumed that only an initial radial gradient was present, which effectively excludes the presence of a significant early classical bulge. The absence of such a classical bulge in the MW is discussed in detail in \S~2.3.1 of \cite{she_zhe_2020}.}. Such a RMG is a natural result if the MW disk formed from the inside-out scenario, which is now well-established \citep{mat_fra_1989, sharma_etal_2021, lu_etal_2022}. \cite{mar_ger_2013} argued that ``mixing during the bar and buckling instabilities is incomplete, and therefore radial metallicity gradients in the initial disk can transform into gradients in the boxy bulge." They also argued that the conservation of Jacobi energy helps stars retain their VMG \citep{mar_ger_2013, bla_ger_2016}. Nevertheless, Jacobi energy is conserved only in a bar rotating with constant pattern speed but not in an evolving potential during the bar and buckling instabilities. Therefore, the detailed formation mechanism of the VMG starting from a RMG still needs further exploration.

\cite{dimatt_etal_2015} systematically discussed the inconsistencies between the model proposed by \cite{mar_ger_2013} and the Galactic observations. They found that in a BPX bulge forming in a cold single-disk model, such as the model of \cite{mar_ger_2013}, the BPX structure is more pronounced in more metal-poor populations. Such a trend contradicts the observations \citep{mcw_zoc_2010, nataf_etal_2010, babusi_etal_2010, ness_etal_2012, catchp_etal_2016}. Besides, they found that the metal-poor population rotates faster than the metal-rich population in such a model, and this trend is also inconsistent with observations. \cite{fragko_etal_2017} compared observations from APOGEE \citep{majews_etal_2017} with two $N$-body models: a single-disk model similar to \cite{mar_ger_2013} (model M2 in their paper) and a triple-disk model (model M1 in their paper). They identified BPX bulge forming in cold single-disk models similar to \cite{mar_ger_2013} also lacks a positive longitudinal metallicity gradient at the disk plane, which is present in both APOGEE and ARGOS data \citep{wylie_etal_2021}. They argued that a single-disk model with initial RMG could not fully explain the Galactic observations and that models with multiple metallicity components similar to their model M1 are necessary to explain the Galactic VMG completely \citep{fragko_etal_2017}. \cite{debatt_etal_2017} proposed the ``kinematic fractionation" scenario, where initially co-spatial disks with different hotness and metallicity can form a VMG, as the disk components with different dynamical hotness form BPX bulges with different spatial extents after secular evolution. Later, \cite{debatt_etal_2020} found that metallicity tagging based on the three initial cylindrical actions rather than the initial radius can improve the match between cold single-disk models and observations.

In this work, we focus on the dynamical origin of the VMG. Instead of attempting to match all known chemical properties of the Galactic Bulge, we restrict our scope to the secular evolution of the Galactic bar, which probably formed more than $8\ \mathrm{Gyr}$ ago \citep{sander_etal_2024}. We construct two $N$-body models, a single-disk model similar to \cite{mar_ger_2013}, and a triple-disk model similar to the model M1 of \cite{fragko_etal_2017}. We confirm that a VMG can form in these secularly evolved BPX models as in \cite{mar_ger_2013}. Using detailed kinematic analyses to trace the dynamical evolution of different stellar populations, we find that the VMG among the two models originates from a ``two-step heating" mechanism driven by the sequential bar and buckling instabilities. We also compare the two models with previous works and find that, despite certain limitations, the ``two-step heating" mechanism is non-negligible in forming the Galactic VMG.

The structure of the paper is organized as follows. In Section \ref{sec: single} and Section \ref{sec: triple}, we explain the setup of the single-disk and triple-disk models and explore the dynamical origin of the vertical metallicity gradient in these models. In Section \ref{sec: dis}, we discuss the implications and limitations of these models. Finally, we summarize the main conclusions in Section \ref{sec: con}.

\section{single-disk Model}\label{sec: single}

\subsection{Model setup}\label{sec: single: model}

\begin{table*}
\caption{Main parameters of the single-disk model}\label{tab: single-disk}
\centering
\begin{tabular}{ccccccccc}
\hline
Component & $M/10^{10} M_\odot$ & $a/\mathrm{kpc}$ & $r_\mathrm{c}/\mathrm{kpc}$ & $\beta$ & $\gamma$ & $N_\mathrm{p}$ \\
DM halo & 70.2 & 11.4 & 150 & 3 & 1 & $\sHaloPartNum$
\\
\hline
Component & $M/10^{10} M_\odot$ & $R_\mathrm{d}/\mathrm{kpc}$ & $h/\mathrm{kpc}$ & $R_{\sigma_R}/\mathrm{kpc}$ & $\sigma_{R,0}/\mathrm{km\cdot s^{-1}}$ & $N_\mathrm{p}$ & $\mathrm{[Fe/H]_0/dex}$ & $\alpha_R/\mathrm{dex\cdot kpc^{-1}}$
\\
Stellar disk & $2.592$ & 1.5 & 0.12 & 3.0 & 150 & $\sDiskPartNum$ & $\sMH0$ & $\sAlphaR0$ \\
\hline
\end{tabular}
\end{table*}

We use \texttt{Agama} \citep{vasili_2019} to construct a $N$-body single-disk galactic model similar to \cite{mar_ger_2013}. The model is a modified version of \cite{tepper_etal_2021}'s, from which we removed the bulge component and rescaled it so that: (1) The final bar radius is approximately 4.5 kpc, similar to the MW \citep{wegg_etal_2015, bla_ger_2016}. (2) Its kinematics roughly match the Galactic observations. 

The model contains a dark matter (DM) halo and a stellar disk. The DM halo has an initial density profile
\begin{equation*}
\rho_\mathrm{s}(r) = \rho_0 (\dfrac{r}{a})^{-\gamma} (1 + \dfrac{r}{a})^{\gamma - \beta} \times \exp{[-(\dfrac{r}{r_\mathrm{c}})^2]},
\end{equation*} where $\rho_0$ is a normalization constant determined by the total mass of the DM halo. It has a density-based quasi-spherical distribution function produced by a generalized Eddington inversion formula implemented by \texttt{Agama} \citep{vasili_2019}. The disk has an initial density profile
\begin{equation*}
\rho_\mathrm{d}(R, z) = \dfrac{\Sigma_0}{4h} \exp{\left(-\dfrac{R}{R_\mathrm{d}}\right)} \times \mathrm{sech}^2{\left(\dfrac{z}{2h}\right)},
\end{equation*} where $\Sigma_0$ is a normalization constant determined by the total mass of the disk. The disk has a quasi-isothermal distribution function:

\begin{equation*}
\begin{aligned}
f(\boldsymbol{J})&=\dfrac{\tilde{\Sigma}\Omega}{2\pi^2 \kappa^2}\times\dfrac{\kappa}{\tilde{\sigma}_R^2}\exp\left(-\dfrac{\kappa J_R}{\tilde{\sigma}_R^2}\right) \\
&\times \dfrac{\nu}{\tilde{\sigma}_Z^2}\exp\left(-\dfrac{\nu J_Z}{\tilde{\sigma}_Z^2}\right)\times\left.\left\{\begin{array}{ll}1&\text{if }J_\phi\geq0,\\\exp\left(\dfrac{2\Omega J_\phi}{\tilde{\sigma}_R^2}\right)&\text{if }J_\phi<0,\end{array}\right.\right.
\end{aligned}
\end{equation*}
where $\kappa$/$\nu$ is the radial/vertical epicycle frequency, $\Omega$ is the circular frequency, $J_R,\ J_\varphi,\ J_Z$ are the actions, and 
\begin{equation*}
\begin{split}
\tilde{\Sigma}(R) &= \Sigma_0 \exp{(-R/R_\mathrm{d})}, \\
\tilde{\sigma}_R^2(R) &= \sigma_{R,0}^2\exp{(-2R/R_{\sigma_R})}, \\
\tilde{\sigma}_Z^2 (R) &= 2 h^2 \nu^2(R).
\end{split}
\end{equation*}
Table \ref{tab: single-disk} lists the main parameters of this model. More details of the model setup are available in \cite{vasili_2019} and \cite{tepper_etal_2021}.

We use \texttt{Gadget4} code \citep{spring_etal_2021} to evolve the model for $\stDuration\ \mathrm{Gyr}$. The model forms a boxy/peanut bulge after the bar and buckling instabilities sequentially occur.

We tag the disk particles with metallicities based on their initial cylindrical radii (similar to \citealt{mar_ger_2013}):
\begin{align}\label{eq: ri based tagging}
\mathrm{[Fe/H]}(R_\mathrm{initial})={\rm[Fe/H]}_0+\alpha_R\times R_\mathrm{initial},
\end{align} 
where $\rm{[Fe/H]}_0=\sMH0\ \mathrm{dex}$ is the initial metallicity at the galactic center, and $\alpha_R=\sAlphaR0\ \mathrm{dex/kpc}$ is the initial RMG (slightly different from those used in \citealt{mar_ger_2013}). The values are selected to achieve a reasonable metallicity range in the precursor disk, of which $\rm{[Fe/H]}_0=\sMH0\ \mathrm{dex}$ is consistent with the ones in the MW (around $-0.3\sim0.6\ \mathrm{dex}$ for stellar populations of different ages in \citealt{lu_etal_2022}), and $\alpha_R=\sAlphaR0\ \mathrm{dex/kpc}$ is consistent with recent observations of high redshift gravitationally lensed galaxies ($-0.30 \sim -0.15\ \mathrm{dex/kpc}$, \citealt{yuan_etal_2011, jones_etal_2013}). Note that we assume no initial VMG in the model, similar to \cite{mar_ger_2013}. Since our focus is Galactic dynamical evolution, we do not consider chemical evolution in this work.

\begin{figure}[htbp!]
\centering
\includegraphics[width=0.5\textwidth, height=0.40\textheight]{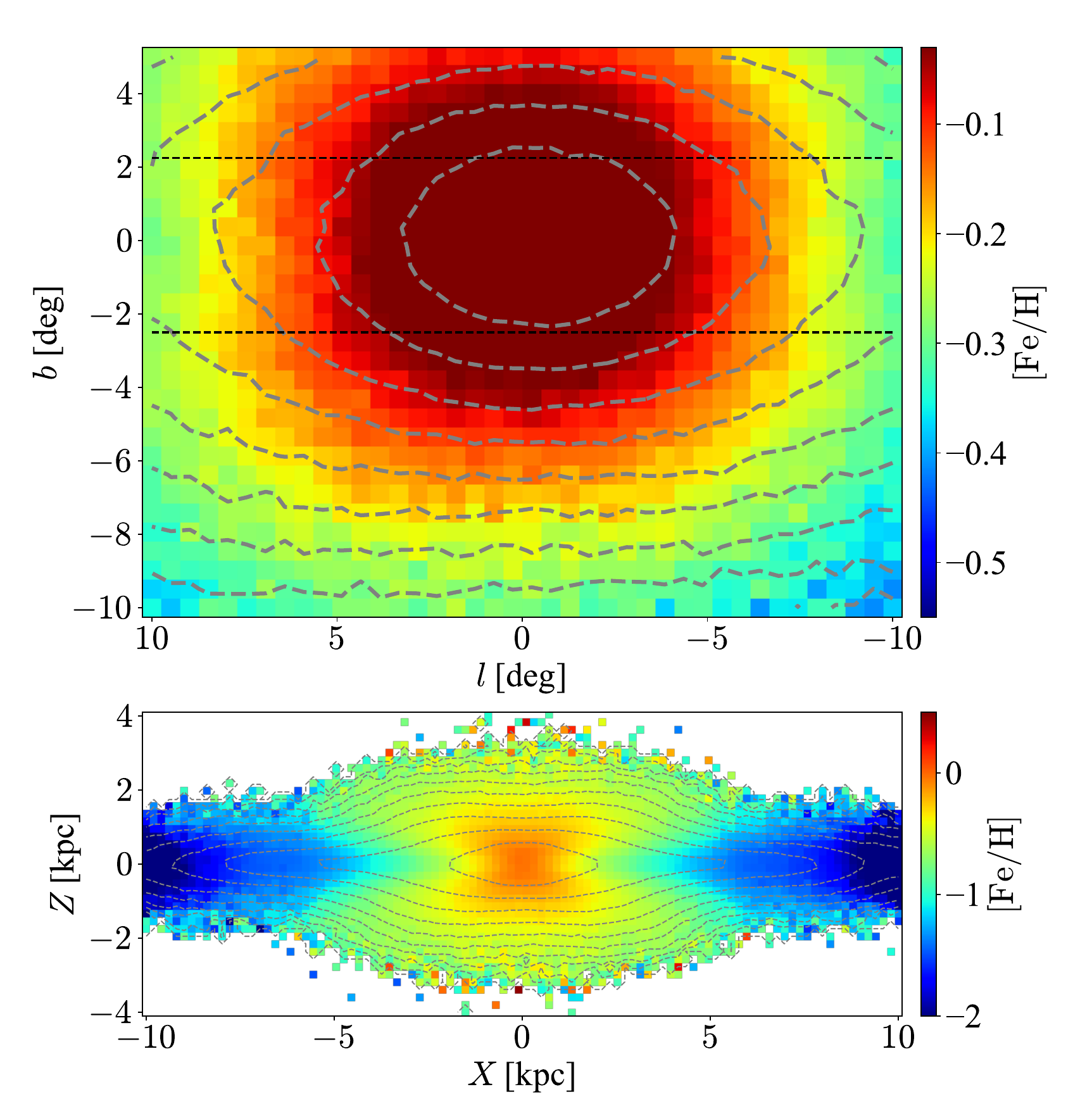}
\caption{The single-disk model's metallicity distribution with grey contours indicates stellar surface density. The top panel shows the distribution in $(l, b)$ coordinates, and the bottom panel displays the side-on distribution (note that the two panels are in different coordinates and spatial ranges). In the top panel, to compare our results with \cite{gonzal_etal_2013}, we limit the range to $-10^\circ\leq b\leq 5^\circ$ and $-10^\circ\leq l\leq 10^\circ$, and we also use black horizontal dashed lines to indicate the region lacking data in \cite{gonzal_etal_2013}. The distribution shows a clear latitudinal gradient at different longitudinal strips.}
\label{fig: single metal map}
\end{figure}

To compare with observations, we set the Galactic Center (GC) at $(X,\ Y,\ Z) = (0,\ 0,\ 0)\ \rm{kpc}$ and the solar position at $(X_\odot,\ Y_\odot,\ Z_\odot) = (-8,\ 0,\ 0.02)\ \rm{kpc}$. We rotate the model so that the bar angle relative to the Sun-GC line is $30^\circ$, closing to the case in the Milky Way \citep{wegg_etal_2015, she_zhe_2020}. In this paper, when we refer to the ``bulge" region, we only consider the particles located between $4\ \rm{kpc}$ and $12\ \rm{kpc}$ from the sun.

\subsection{VMG in the single-disk model}\label{sec: single: VMG}

Figure \ref{fig: single metal map} shows the metallicity distribution of the bulge in the single-disk model. The top panel shows that the single-disk model exhibits a VMG pattern akin to those reported by \cite{mar_ger_2013} and \cite{gonzal_etal_2013}. The VMG of the single-disk model along the bulge minor axis is about $\sMinorVMG\ \mathrm{dex/deg}$. It is calculated by linearly fitting pixels that $|l|\leq0.5^\circ$ and $-10^\circ\leq b\leq-3^\circ$ in the top panel of Figure \ref{fig: single metal map}, from which the Pearson correlation coefficient is less than $-0.9$. Such a value closely resembles the observed value of the MW, which is approximately $-0.04\ \rm{dex/deg}$ \citep{gonzal_etal_2013}. Therefore, the single-disk model successfully confirms the findings of \cite{mar_ger_2013} and is consistent with the VMG pattern observed in the MW. This result demonstrates that the secular evolution of a disk can qualitatively reproduce a VMG.

\begin{figure*}[htbp!]
\centering
\includegraphics[width=1.\textwidth, height=.6\textheight]{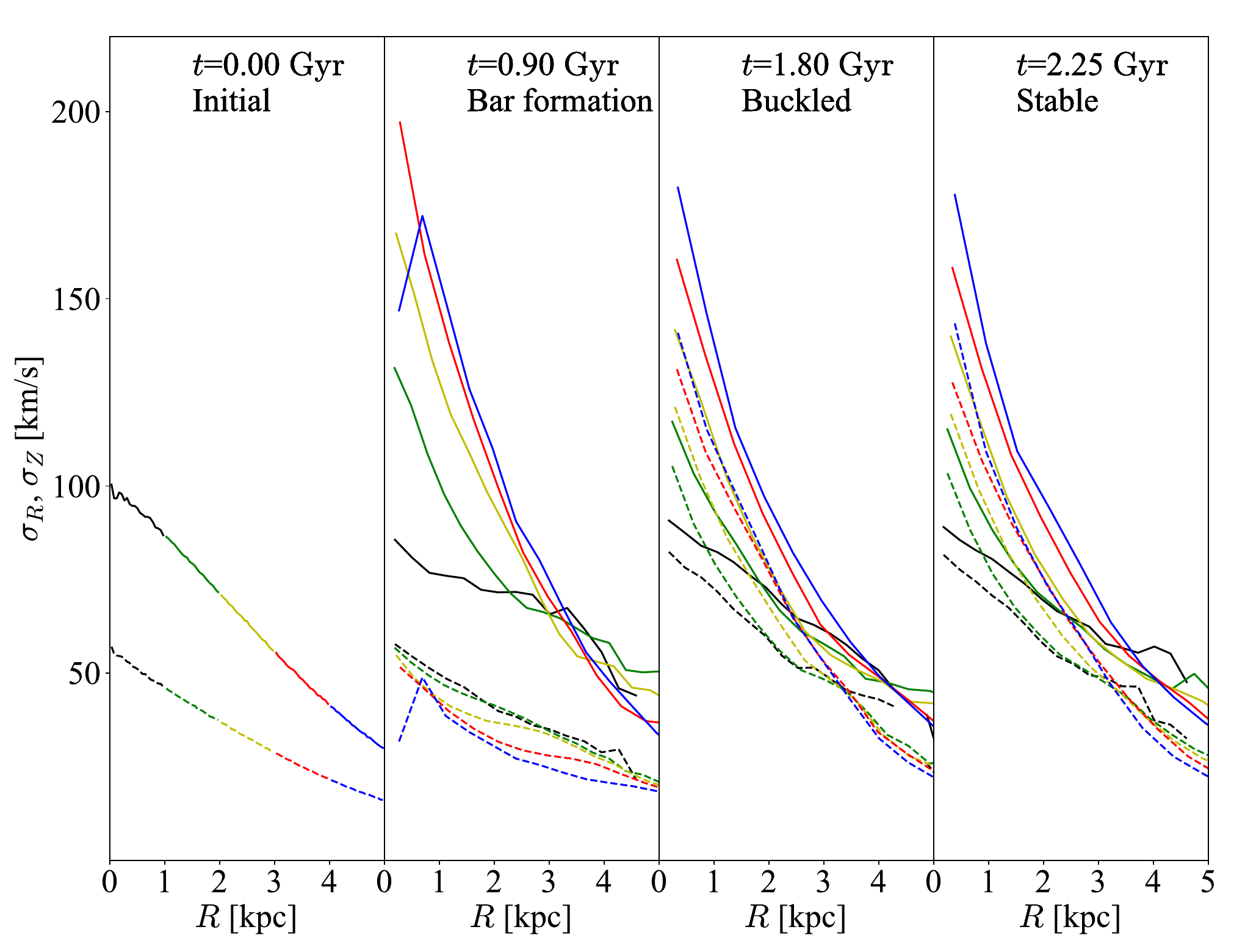}
\caption{The cylindrical radial (solid lines) and vertical (dashed lines) velocity dispersion profiles of the single-disk model. The epoch of each panel from left to right: initial snapshot at $0.00\ \rm{Gyr}$, bar formed but before buckling at $0.90\ \rm{Gyr}$, near the saturation of buckling at $1.80\ \rm{Gyr}$ and a snapshot in quasi-steady state at $2.25\ \rm{Gyr}$. Particles in different 1 $\rm{kpc}$ initial radii bins are color-coded with different colors, from inside to outside: black, green, yellow, red, and blue. The dispersion profiles vary significantly from the first column to the second and from the second to the third, indicating violent radial heating and vertical heating during bar and buckling instabilities, respectively; see more details in the context.}
\label{fig: single two-step}
\end{figure*}

\begin{figure*}[htbp!]
\centering
\includegraphics[width=.495\textwidth, height=.28\textheight]{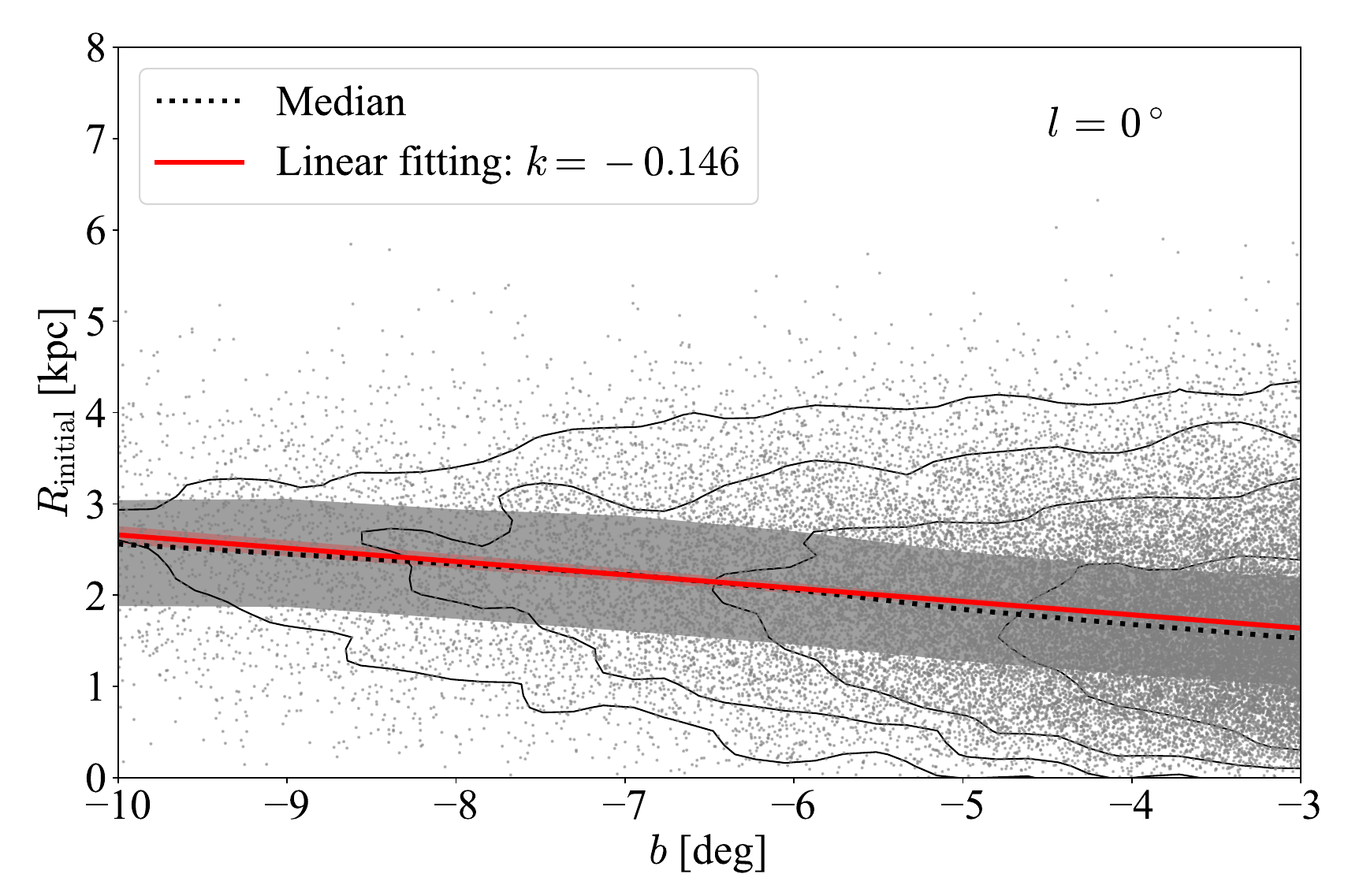}
\includegraphics[width=.495\textwidth, height=.28\textheight]{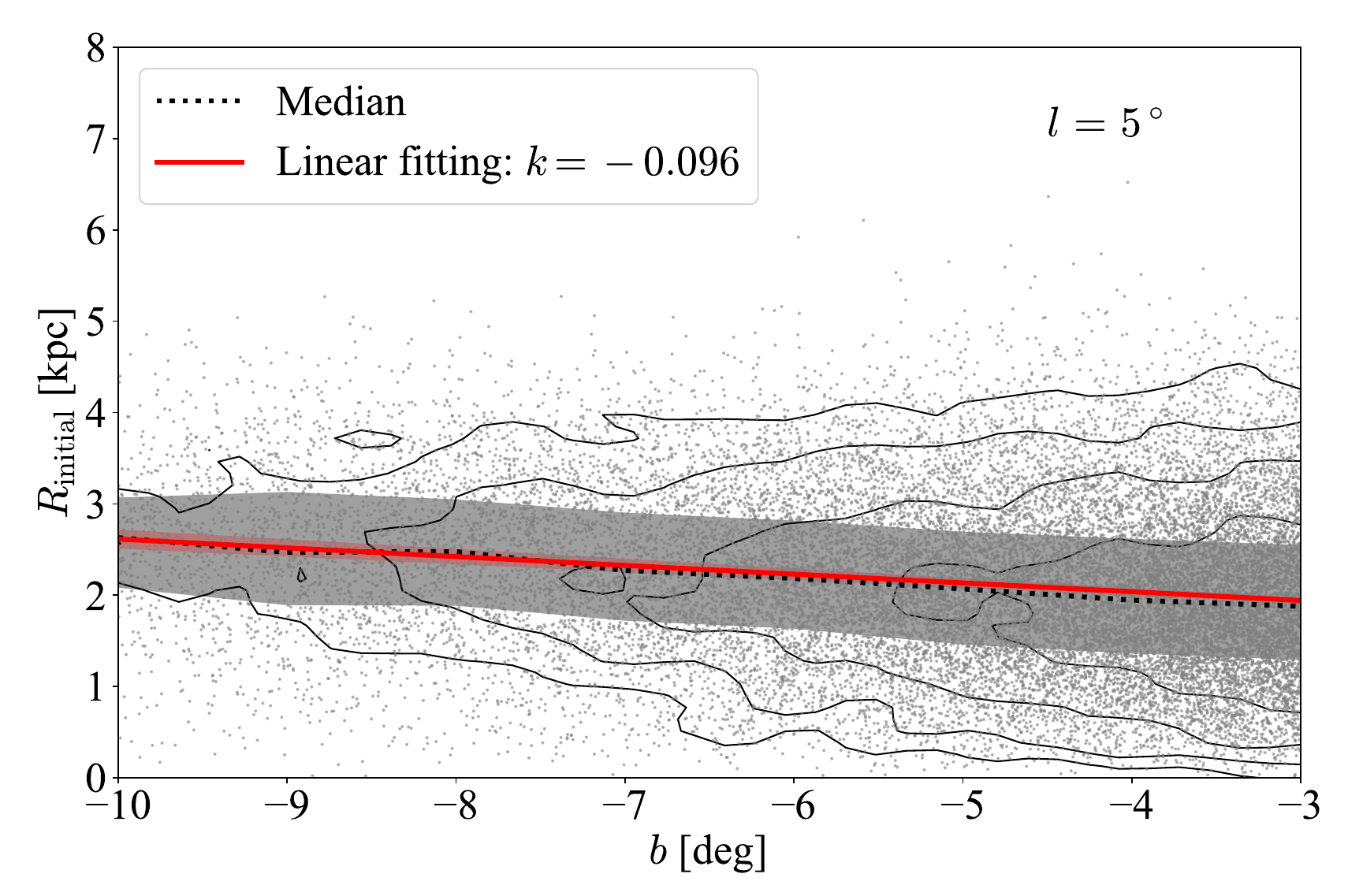}
\includegraphics[width=.495\textwidth, height=.28\textheight]{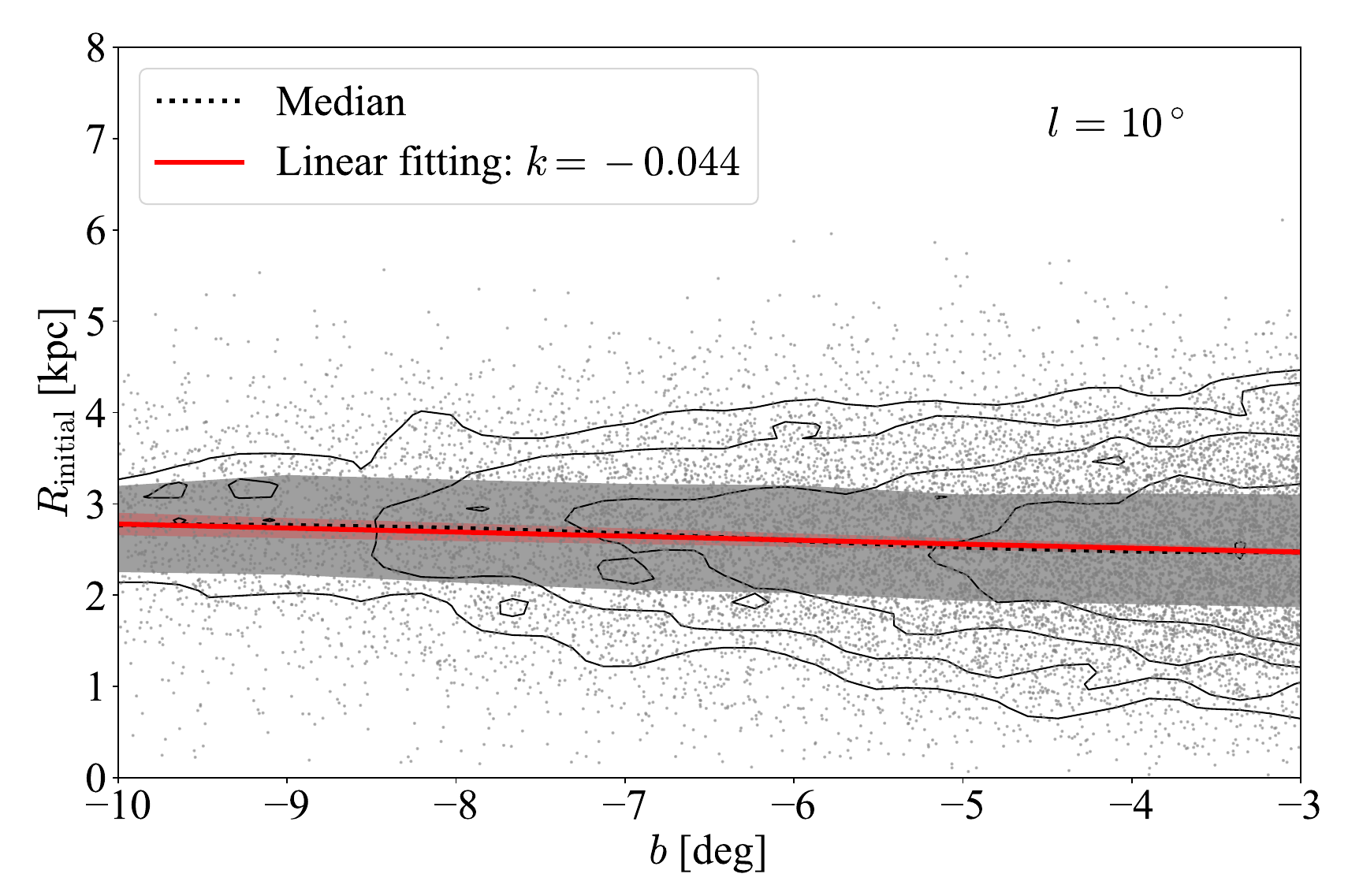}
\includegraphics[width=.495\textwidth, height=.25\textheight]{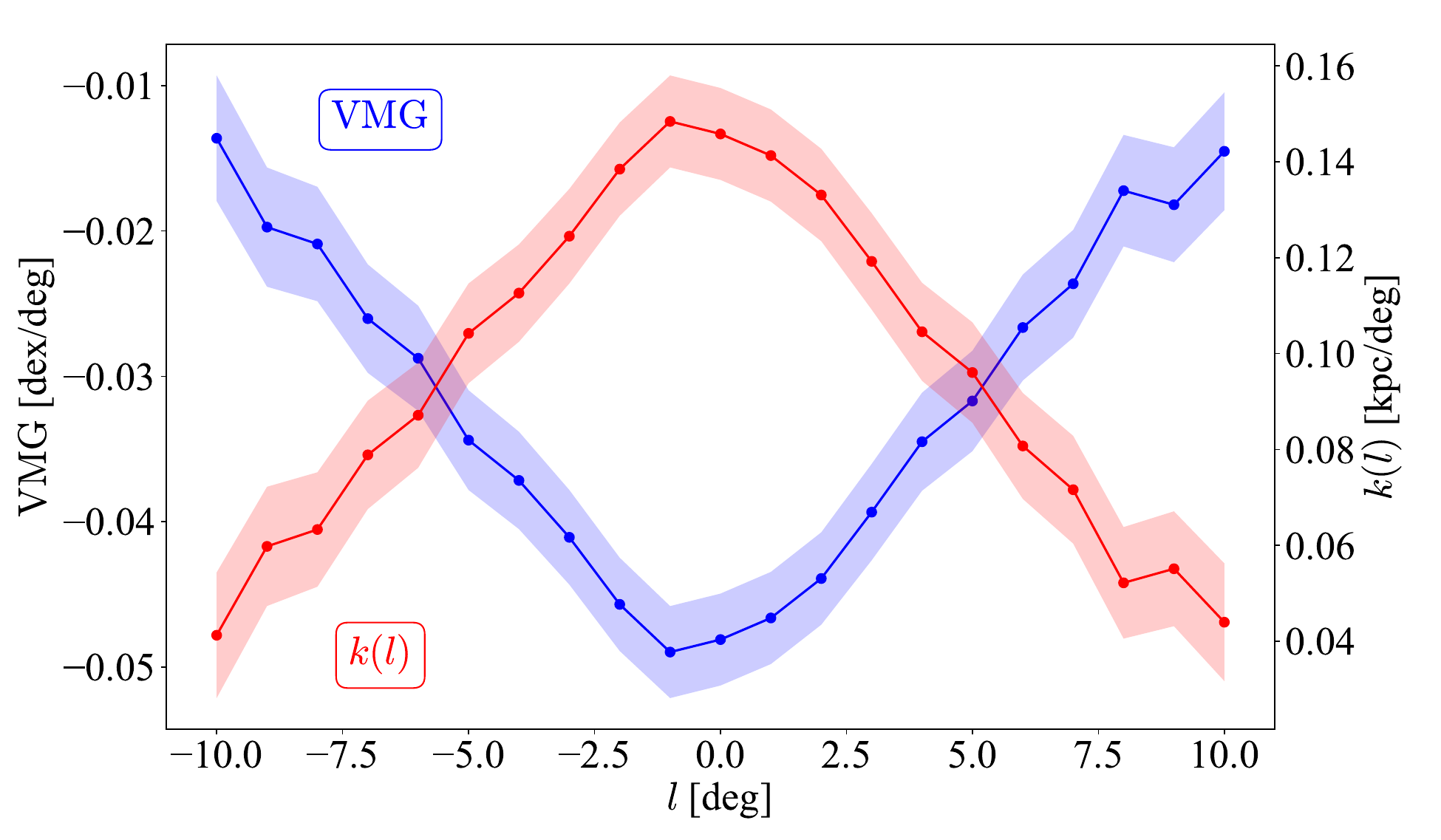}
\caption{The top left three panels display the $b_\mathrm{final}-R_\mathrm{initial}$ distribution of bulge particles in the single-disk model across three $1^\circ$ width longitudinal strips at $l=0^\circ$, $5^\circ$, and $10^\circ$. In each panel, black contours represent the number density of particles. Black dotted lines show the median values within $\Delta b=1^\circ$ bins, with gray shadows indicating the higher-lower quartile. Red solid lines depict the linear fittings of the distributions, which are calculated with \texttt{scipy.stats.linregress}, and their surrounding red shadows display the three-times standard errors of the slopes. The bottom right panel illustrates how $\rm{VMG}$ (red) and $k(l)$ (blue) vary with $l$. $\mathrm{VMG}$/$k(l)$ is calculated from a linear fit of $\rm{[Fe/H]}$/$R_\mathrm{initial}$ versus $|b|$ at different $l$, using bulge particles within $-10^\circ\leq b\leq-3^\circ$. Again, the shadows represent the three-times standard errors of the fitting slopes. Notably, the $\mathrm{VMG}$ and $k(l)$ are mirror symmetric in the bottom right panel, consistent with the expectation from the Equation \ref{eq: single-disk formula}.}
\label{fig: single ri vs. bf}
\end{figure*}

\subsection{Dynamical origin of the VMG: a ``two-step heating" mechanism}\label{sec: single: mechanism}

Equation \ref{eq: ri based tagging} shows that particles with lower metallicity are initially located in the outer part of the disk. Therefore, the VMG present in Figure \ref{fig: single metal map} implies the existence of radial and vertical mixing processes of particles. To investigate these processes in detail, we present the model's radial and vertical dispersion profiles in Figure \ref{fig: single two-step} at four epochs: the initial state, a pre-buckling bar epoch, an epoch near the saturation of buckling, and an epoch in a quasi-steady secular evolution state. Figure \ref{fig: single two-step} illustrates how the radial and vertical mixing processes in the single-disk model work together to generate a VMG, which we name a ``two-step heating" mechanism.

The first step is radial heating due to the bar instability. Initially, both radial and vertical dispersions exhibit nearly exponential profiles. As time progresses, the onset of bar instability (between the first and second epochs) redistributes particles across a broader radial range and heats them (see the second column of Figure \ref{fig: single two-step}). The heating effect is more pronounced in the radial direction than the vertical direction, owing to the in-plane nature of the bar instability. Consequently, bar formation increases the $\sigma_R/\sigma_Z$ ratio at all radii in the disk. The $\sigma_R/\sigma_Z$ ratio rises significantly in the particles with larger initial radii, as their lower initial dispersion (see the first column of Figure \ref{fig: single two-step}) makes them more susceptible to heating processes during bar formation. Therefore, the more metal-poor populations gain higher $\sigma_R/\sigma_Z$ ratios after the first step.

The second step is vertical heating due to the buckling instability. After the first step, the increased $\sigma_R/\sigma_Z$ ratio of the disk triggers the buckling instability \citep{raha_etal_1991, mer_sel_1994}. During the buckling, particles are vertically mixed and heated (see the third column of Figure \ref{fig: single two-step}). Since more metal-poor populations have a higher $\sigma_R/\sigma_Z$ ratio after the first step, they undergo a more robust buckling process and, consequently, a more violent vertical heating process. Therefore, more metal-poor particles end up at greater vertical heights after buckling, naturally giving rise to a VMG.

In summary, the VMG naturally develops through a ``two-step heating" mechanism: the onset of bar instability induces radial mixing and heating, and subsequent buckling instability triggers vertical mixing and heating. The more metal-poor particles, which are initially outer and so colder, undergo more violent heating processes during the sequential bar and buckling instabilities and move to higher vertical heights.

\subsection{Relationship between the VMG and the initial RMG}\label{sec: single: formula}

To understand how the ``two-step heating" mechanism transforms the initial RMG into the VMG in the single-disk model, we display the final distribution of bulge particles' initial radii versus latitudes in three longitudinal strips in the first three panels of Figure \ref{fig: single ri vs. bf}. In each panel, the medians of initial radii are nearly linearly correlated with latitudes, as indicated by the consistency between the black dotted lines for medians and the red solid lines for the linear fitting results. Thus, the ``two-step heating" mechanism redistributes the particles nearly linearly in the latitudinal direction. Such a nearly linear redistribution is also present in \cite{debatt_etal_2020} (see their Figure 9), except that they considered $R_\mathrm{initial}$ versus $Z$ rather than $|b|$, and consider all stellar particles in a greater vertical extent. Hence, when the particles are assigned with metallicity according to Equation \ref{eq: ri based tagging}, the linearly distributed $R_\mathrm{initial}$ is mapped into the VMG in Figure \ref{fig: single metal map} as follows:

\begin{equation} \label{eq: single-disk formula}
\begin{split}
&\mathrm{VMG}(l)\equiv \dfrac{\partial \left<\mathrm{[Fe/H]}\right>}{\partial |b_\mathrm{final}|}\bigg|_l\\
&= \dfrac{\partial \left<\mathrm{[Fe/H]}_0 + \alpha_R\times R_\mathrm{initial}\right>}{\partial |b_\mathrm{final}|}\bigg|_l\\
&= \alpha_R \dfrac{\partial\left<R_\mathrm{initial}\right>}{\partial |b_\mathrm{final}|}\bigg|_l \\
&= \alpha_R\,k(l),\ (\mathrm{where}\ k(l)\equiv\dfrac{\partial\left<R_\mathrm{initial}\right>}{\partial |b_\mathrm{final}|}\bigg|_l),
\end{split}
\end{equation}
where the factors $k(l)=\dfrac{\partial\left<R_\mathrm{initial}\right>}{\partial |b_\mathrm{final}|}\bigg|_l$ can be determined by a linear fitting of $R_\mathrm{initial}$ versus $|b_\mathrm{final}|$ for bulge particles at different longitudes. Naturally, the $k(l)$ is already determined for a specific snapshot. In this context, the VMG in such a snapshot is directly proportional to the initial RMG $\alpha_R$. This expectation is verified by the reflection symmetry of VMGs and $k(l)$ factors in the bottom right panel of Figure \ref{fig: single ri vs. bf}, where we use the bulge particles within $-10^\circ\leq b\leq -3^\circ$ to calculate the VMGs and $k(l)$ factors. 

Notably, the $k(l)$ factor depends on the galactic evolutionary history: In the bottom right panel of Figure \ref{fig: single ri vs. bf}, the $\mathrm{VMG}$ exhibits an extreme near the minor axis and gradually flattens away from it. This trend is a natural result of the ``two-step heating" mechanism, among which the radial and vertical heating processes are more violent near the GC during the ``two-step heating" processes (see Figure \ref{fig: single two-step}). Moreover, the bottom right panel of Figure \ref{fig: single ri vs. bf} shows a slightly steeper $\rm{VMG}$ on the positive longitude side than the negative side because the positive side is closer to the GC than the negative side. So, particles on the positive side undergo a more violent heating process. In conclusion, a more violent heating history in the ``two-step heating" processes should lead to a higher $k(l)$ value and, consequently, a steeper VMG.

As a byproduct of the transformation from the RMG to the VMG, the RMG in the disk is partially flattened by the radial mixing process, resulting in a final RMG approximately $\sFinalRMG\ \mathrm{dex/kpc}$ on the disk within $10\ \rm{kpc}$ from the GC.

\begin{table*}
\caption{Main parameters of the triple-disk model}\label{tab: ri based triple-disk}
\centering
\begin{tabular}{cccccccccc}
\hline
Component & $M/10^{10} M_\odot$ & $a/\mathrm{kpc}$ & $r_c/\mathrm{kpc}$ & $\beta$ & $\gamma$ & $N_\mathrm{p}$ \\
DM halo & 70.875 & 15 & 150 & 3 & 1 & $\tHaloPartNum$ \\
\hline
Component & $M/10^{10} M_\odot$ & $R_\mathrm{d}/\mathrm{kpc}$ & $h/\mathrm{kpc}$ & $R_{\sigma_R}/\mathrm{kpc}$ & $\sigma_{R,0}/\mathrm{km\cdot s^{-1}}$ & $N_\mathrm{p}$ & $\mathrm{[Fe/H]_0/dex}$ & $\alpha_R/\mathrm{dex\cdot kpc^{-1}}$ \\
Thin disk & 1.705 & 2.4 & 0.075& 4.8 & 90 & $\tThinPartNum$ & $\thinMH0$ & $\thinAlphaR0$ \\
Intermediate disk & 1.041 & 1.0 & 0.2 & 2.0 & 117 & $\tInterPartNum$ & $\interMH0$ & $\interAlphaR0$\\
Thick disk & 0.753 & 1.0 & 0.3 & 2.0 & 144 & $\tThickPartNum$ & $\thickMH0$ & $\thickAlphaR0$ \\
\hline\noalign{\smallskip}
\end{tabular}
\end{table*}

\section{Triple-disk model}\label{sec: triple}

\subsection{Model setup}\label{sec: triple: model}

To verify the ``two-step heating" mechanism in more complex cases, we construct a triple-disk model similar to the model M1 of \cite{fragko_etal_2017}. Each disk in this model has a configuration similar to the disk of the single-disk model, and their main parameters are listed in Table \ref{tab: ri based triple-disk}. In this model, we call the three disks the thin, intermediate, and thick disks according to their scale heights. We use the \texttt{Gadget4} code \citep{spring_etal_2021} to evolve this model for $\ttDuration\ \mathrm{Gyr}$. This model also forms a boxy/peanut bulge after the bar and buckling instabilities sequentially occur.

Again, we rotate the model so that the bar angle relative to the Sun-GC line is $30^\circ$. We replace the three radially independent Gaussian metallicity distributions in the model M1 of \cite{fragko_etal_2017} with three initial RMGs similar to the single-disk model (see Equation \ref{eq: ri based tagging}) in this section and delay the discussion of their approach to \S~\ref{sec: disc: 3DG}. The parameters ($\mathrm{[Fe/H]}_0$,\ $\alpha_R$) of each disk are also listed in Table \ref{tab: ri based triple-disk}.

\begin{figure*}[htbp!]
\centering
\includegraphics[width=0.45\textwidth, height=0.36\textheight]{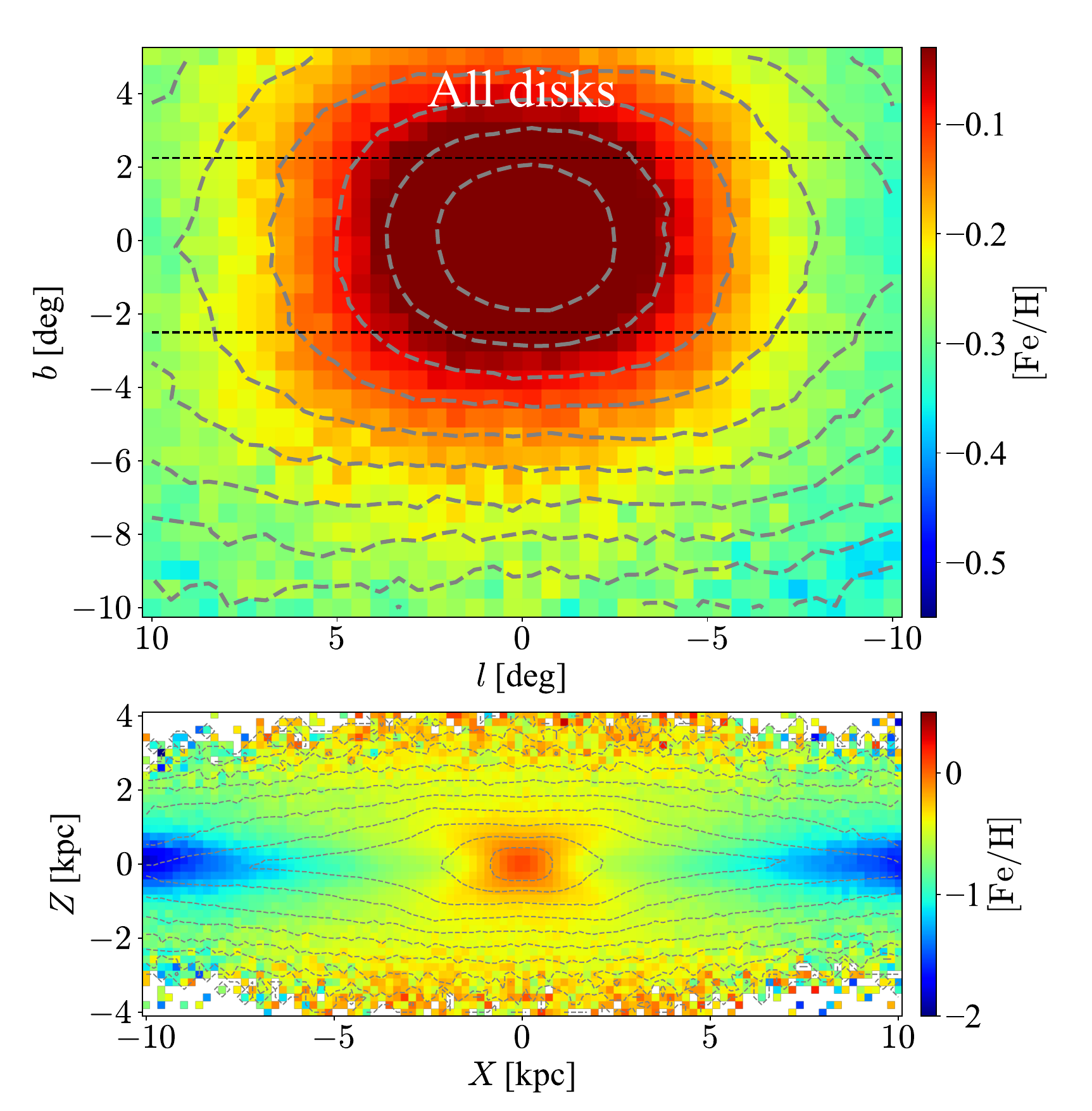}
\includegraphics[width=0.45\textwidth, height=0.36\textheight]{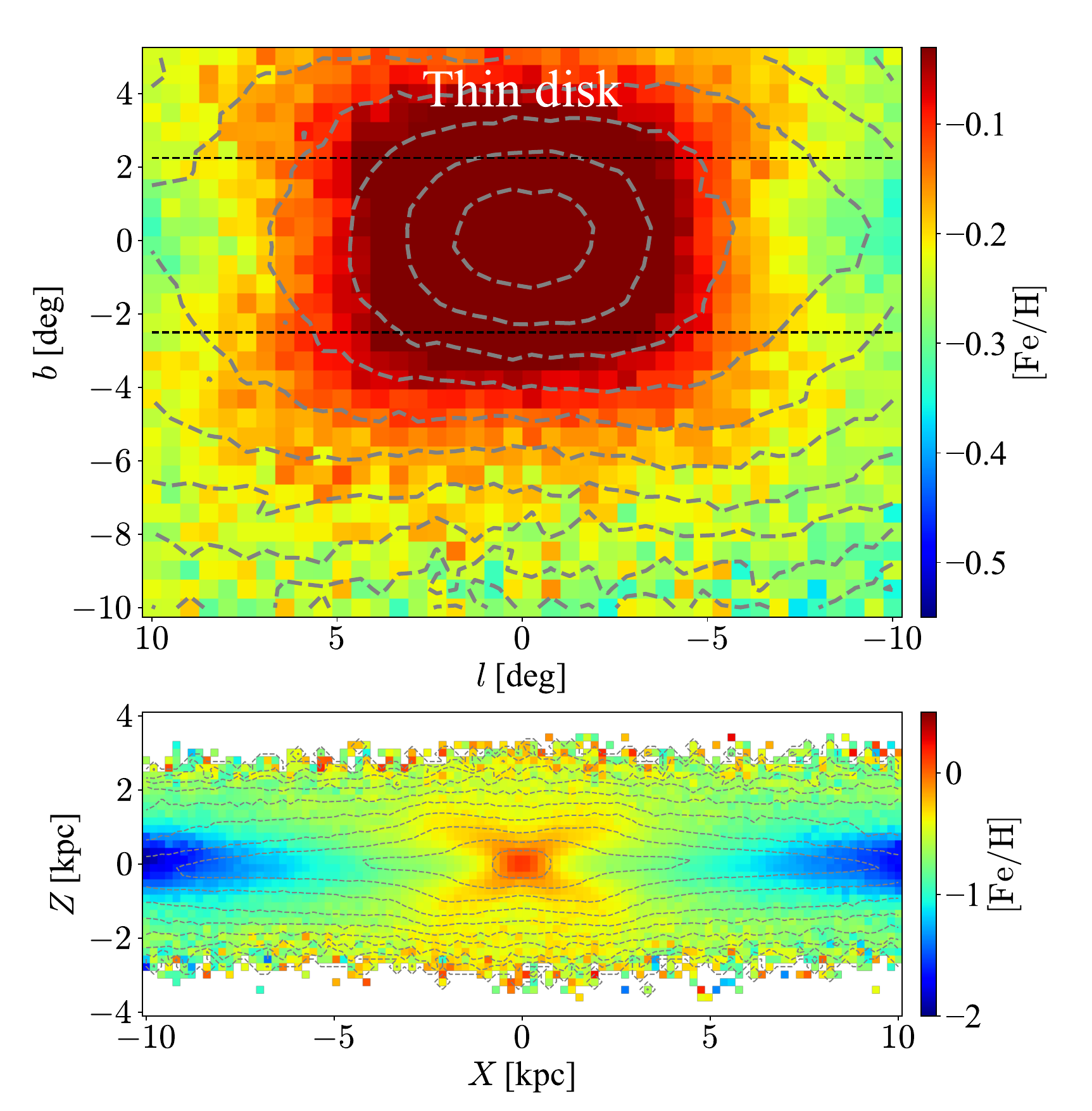}
\includegraphics[width=0.45\textwidth, height=0.36\textheight]{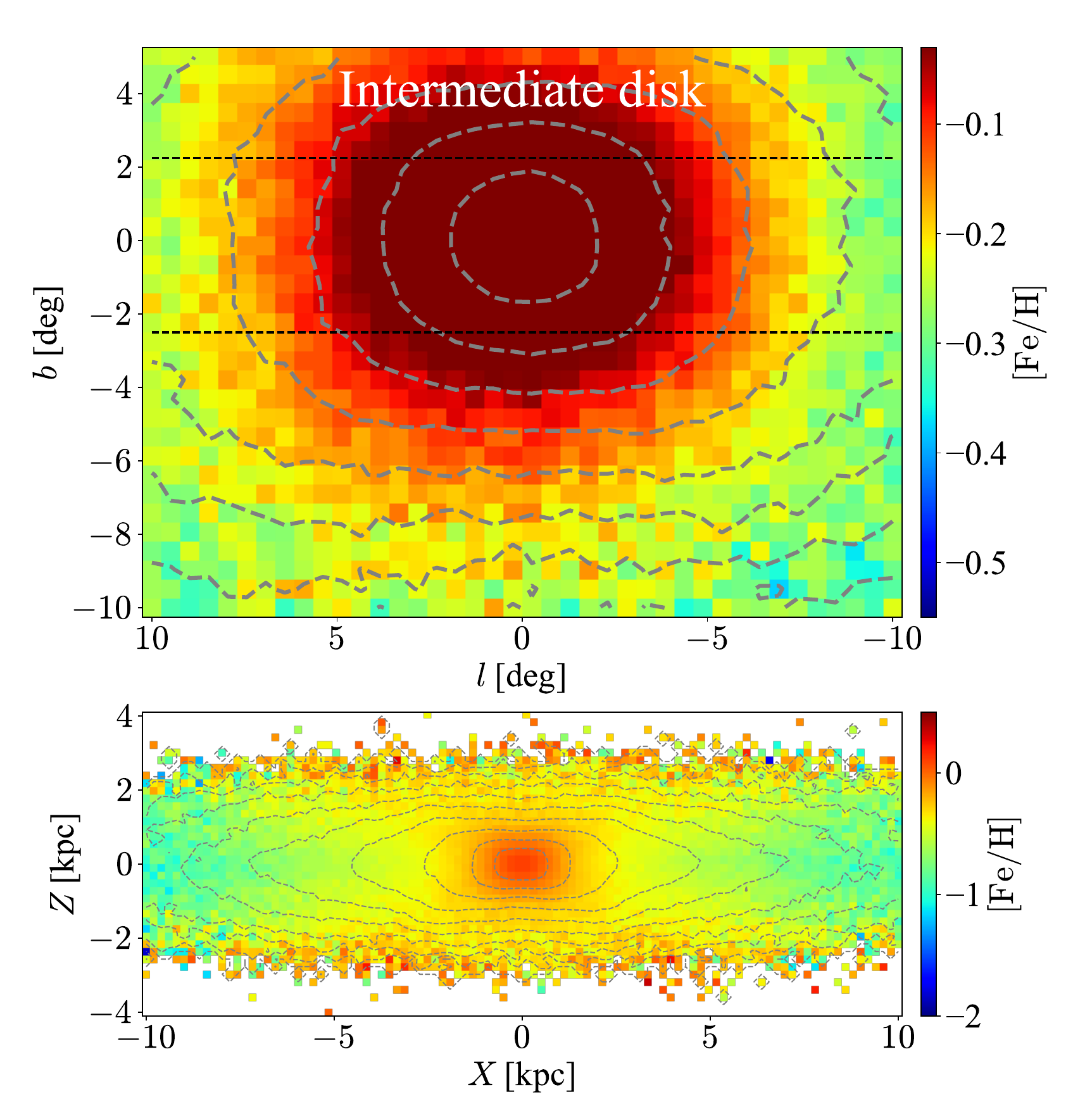}
\includegraphics[width=0.45\textwidth, height=0.36\textheight]{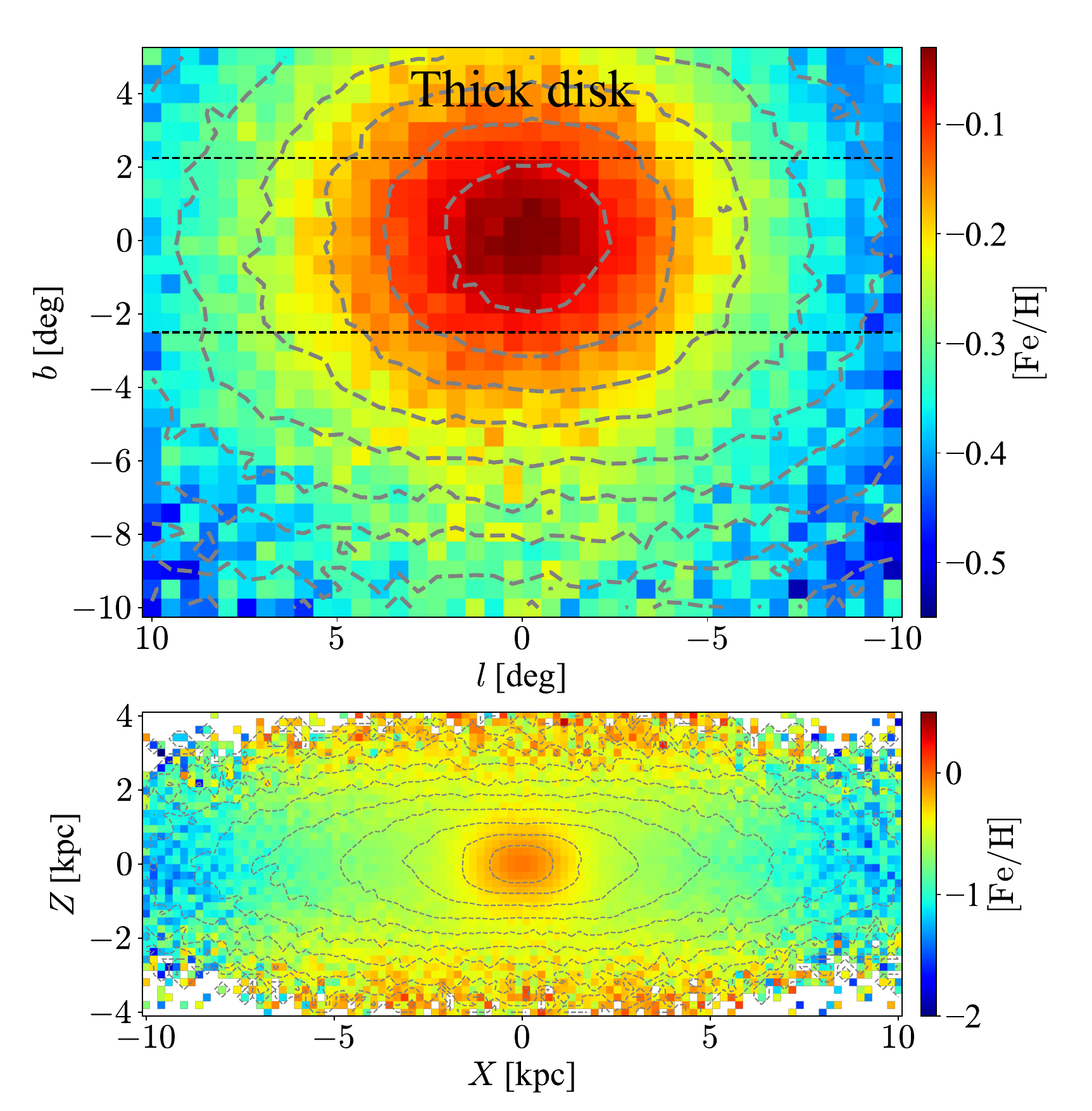}
\caption{Similar to Figure \ref{fig: single metal map}, but for the triple-disk model and its three disk components. The distributions resemble the one in the single-disk model.}
\label{fig: triple ri tag metal map}
\end{figure*}

\subsection{VMG in the triple-disk model}\label{sec: triple: VMG}

Figure \ref{fig: triple ri tag metal map} shows the bulge metallicity distribution of the triple-disk model for both the entire model and its three disk components. Qualitatively, the global trends of the metallicity distribution in the four panels mirror the distribution of the single-disk model. The triple-disk model and all its disk components show a VMG across the bulge region associated with a metal-rich core, similar to the single-disk model. The VMG at the bulge minor axis in the triple-disk model is about $\tMinorVMG\ \mathrm{dex/deg}$, calculated by linearly fitting pixels that $|l|\leq0.5^\circ$ and $-10^\circ\leq b\leq-3^\circ$ in the top left panel of Figure \ref{fig: triple ri tag metal map}. Such a VMG also resembles the Galactic observations ($-0.04\ \mathrm{dex/deg}$ in \citealt{gonzal_etal_2013}), confirming the secular evolution can also qualitatively reproduce a VMG similar to the MW in a multiple-disk model.

\begin{figure*}[htbp!]
\centering
\includegraphics[width=.495\textwidth, height=0.285\textheight]{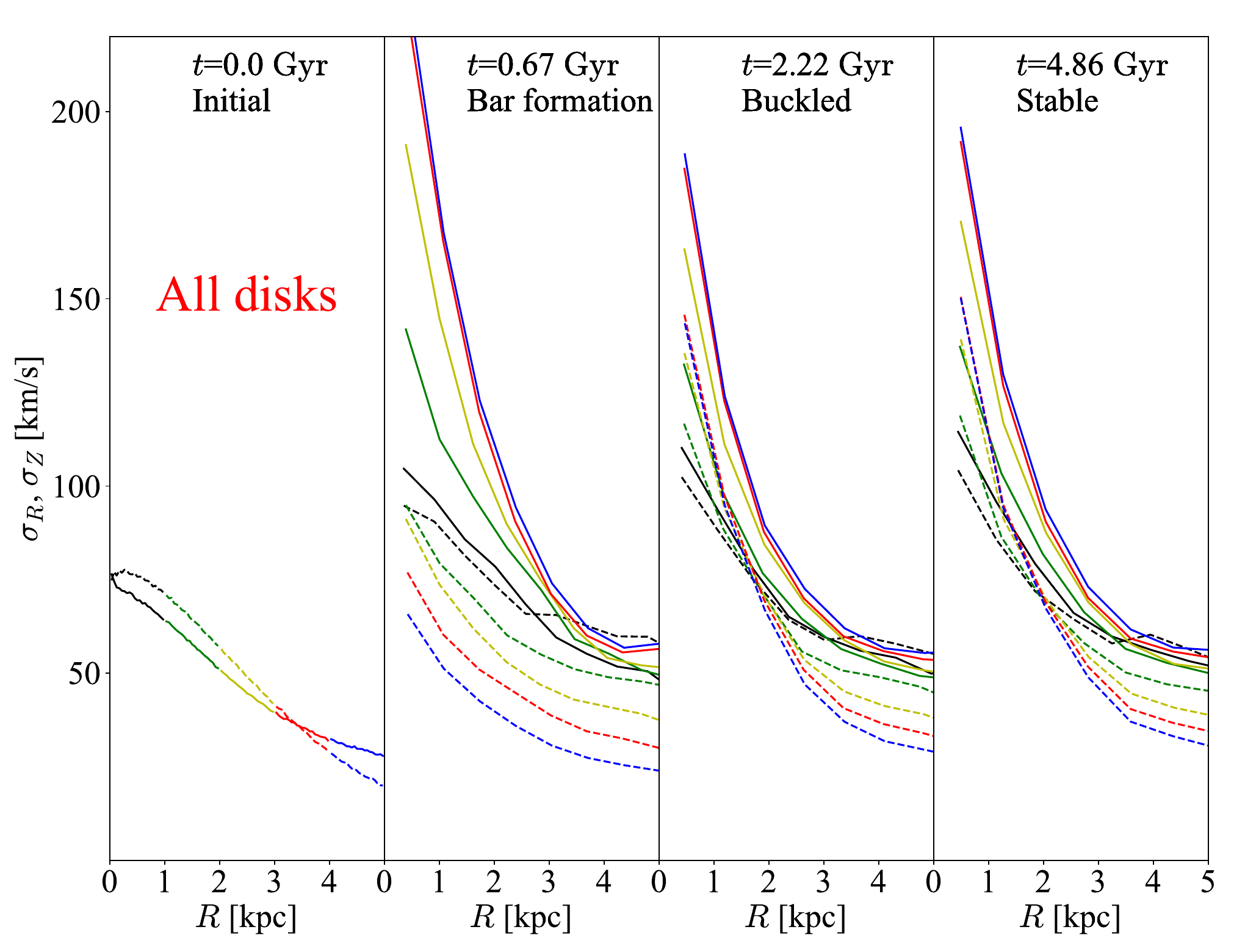}
\includegraphics[width=.495\textwidth, height=0.285\textheight]{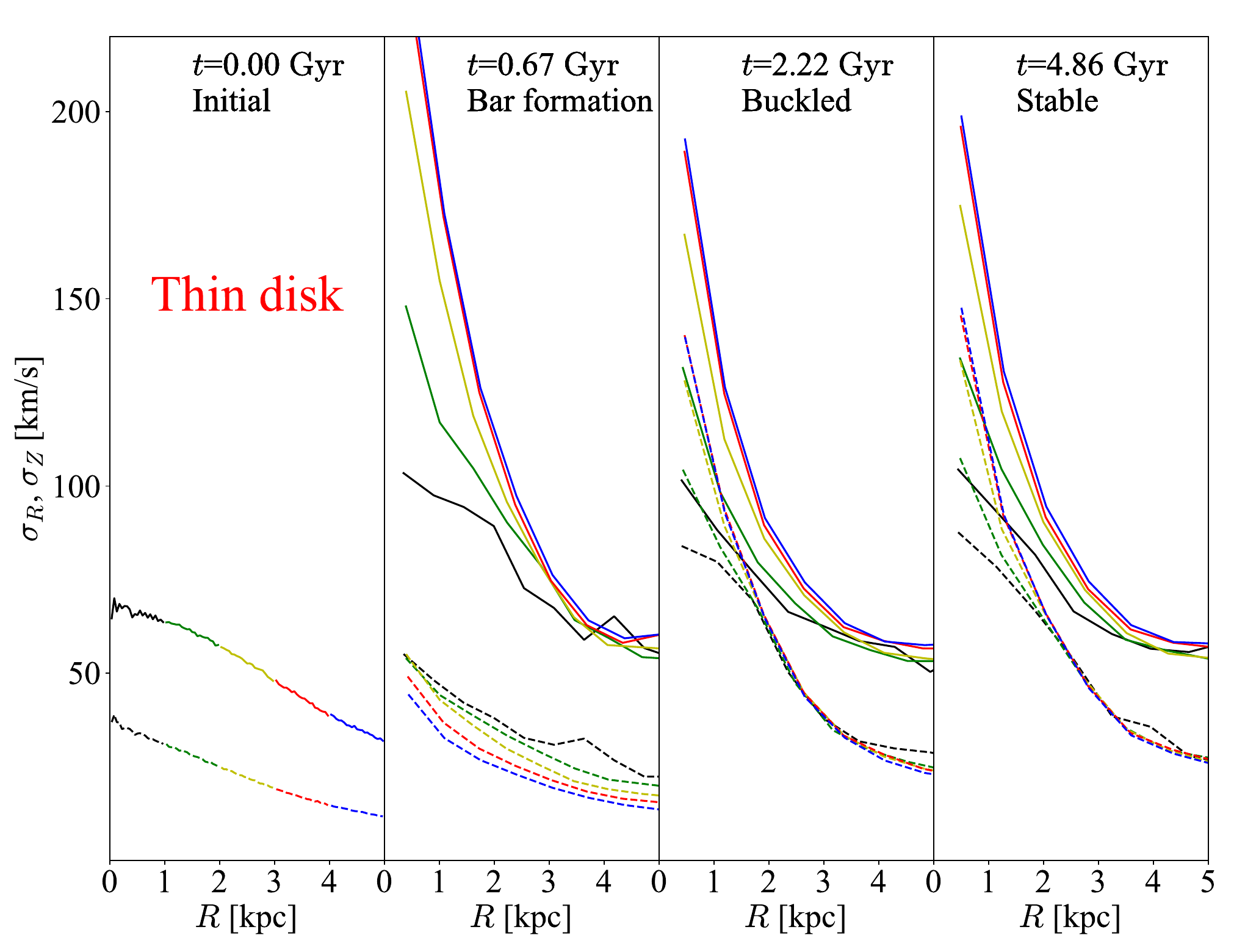}
\includegraphics[width=.495\textwidth, height=0.285\textheight]{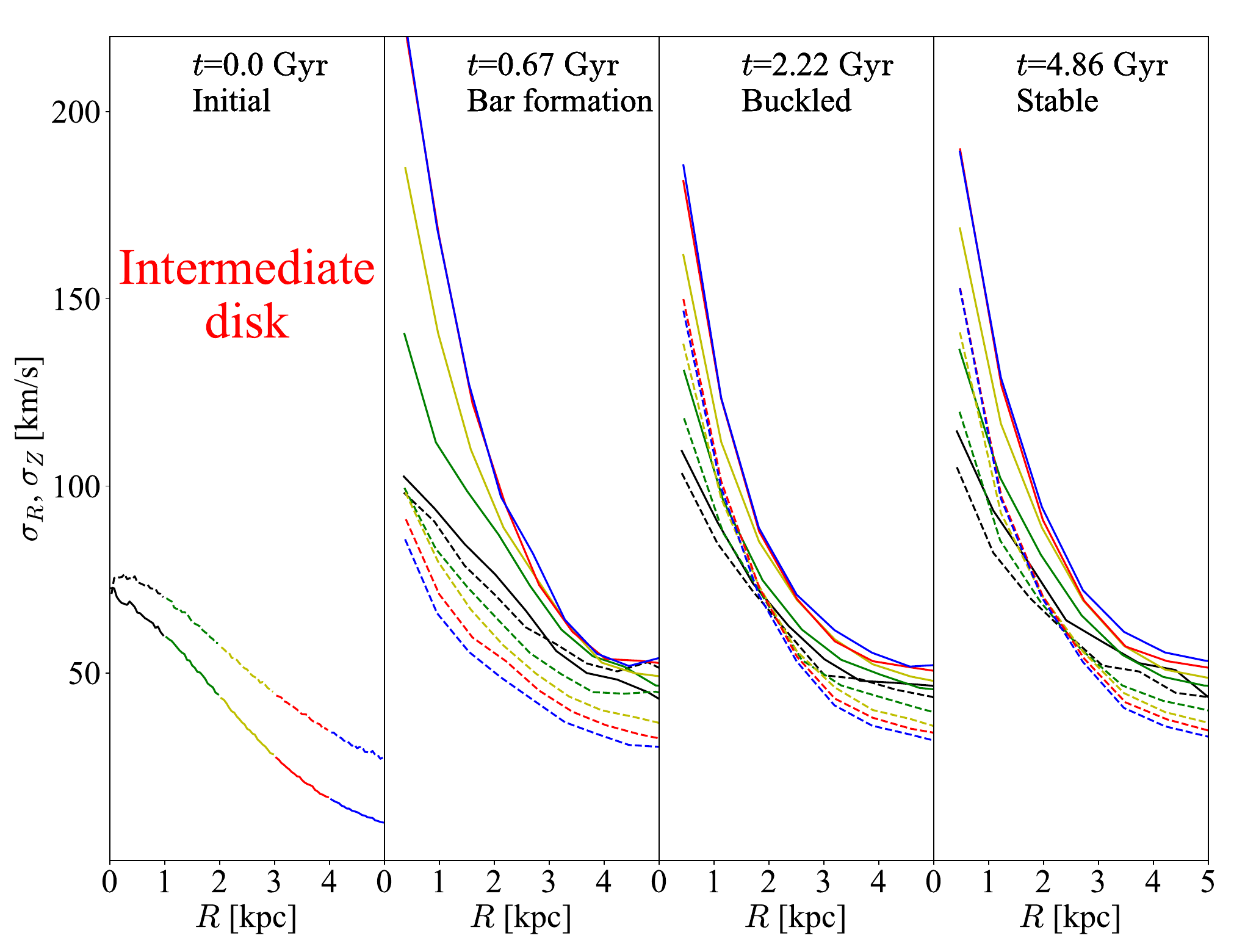}
\includegraphics[width=.495\textwidth, height=0.285\textheight]{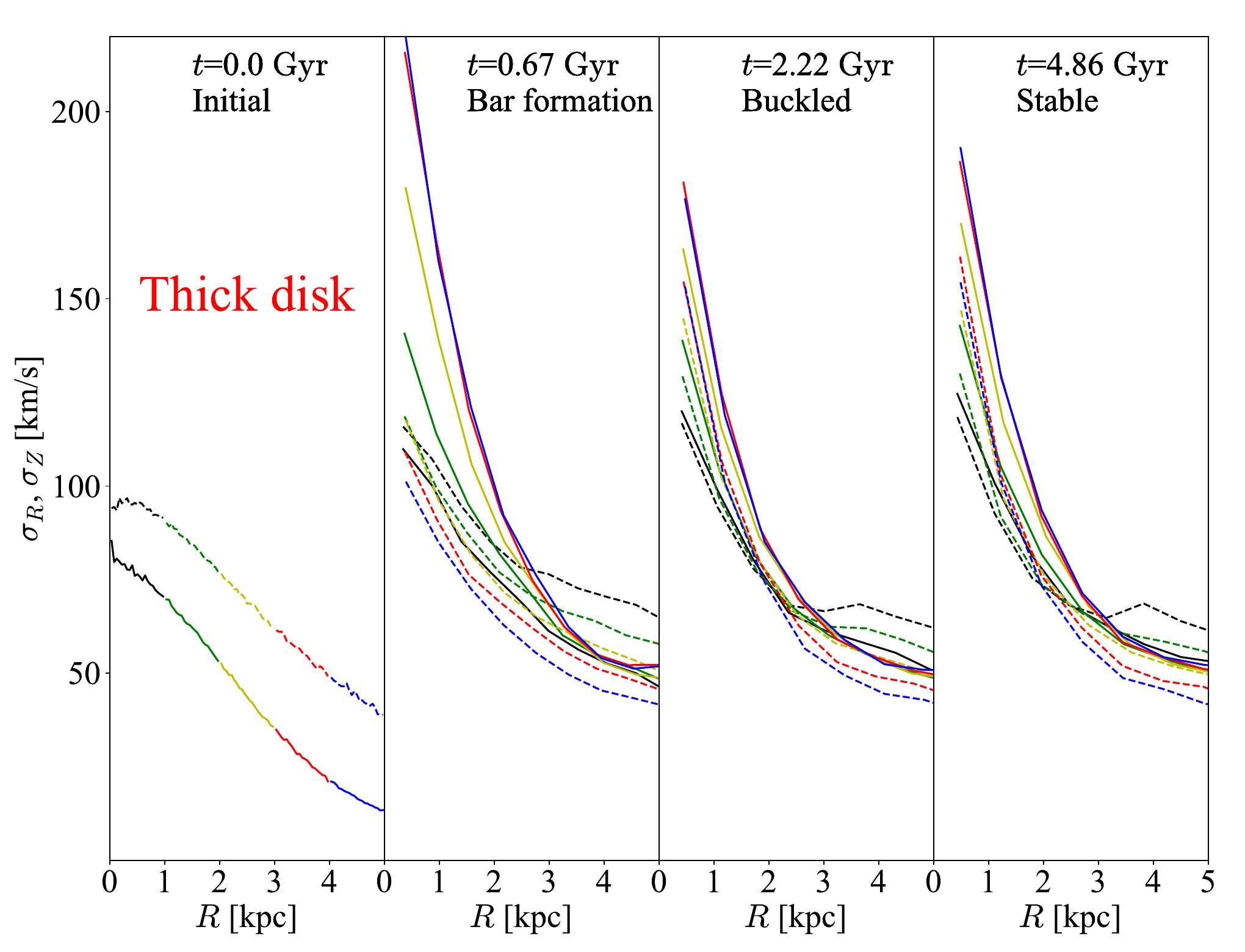}
\caption{Top left panel: The cylindrical radial (solid lines) and vertical (dashed lines) velocity dispersion profiles in the triple-disk model, similar to Figure \ref{fig: single two-step}. The other three panels display the velocity dispersion profiles for individual disk components: the thin disk (top right), the intermediate disk (lower left), and the thick disk (lower right). The triple-disk model exhibits an evolutionary history similar to the single-disk model shown in Figure \ref{fig: single two-step} for both the entire model and each disk component. They all undergo pronounced radial heating during the bar instability and vertical heating during the buckling instability. Therefore, the ``two-step heating" mechanism also operates in the triple-disk model.}
\label{fig: triple two-step}
\end{figure*}

\begin{figure*}[htbp!]
\centering
\includegraphics[width=.495\textwidth, height=.28\textheight]{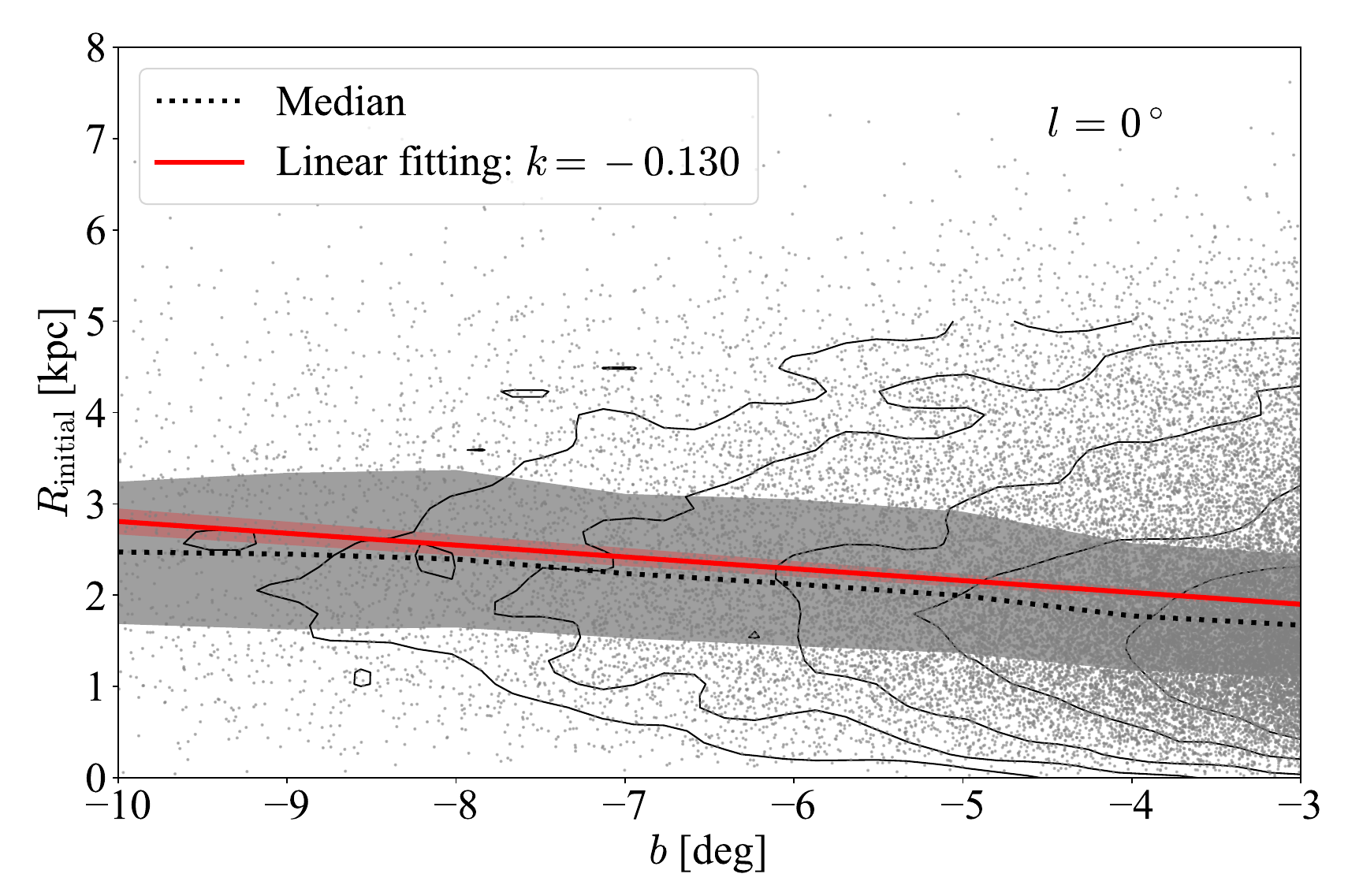}
\includegraphics[width=.495\textwidth, height=.28\textheight]{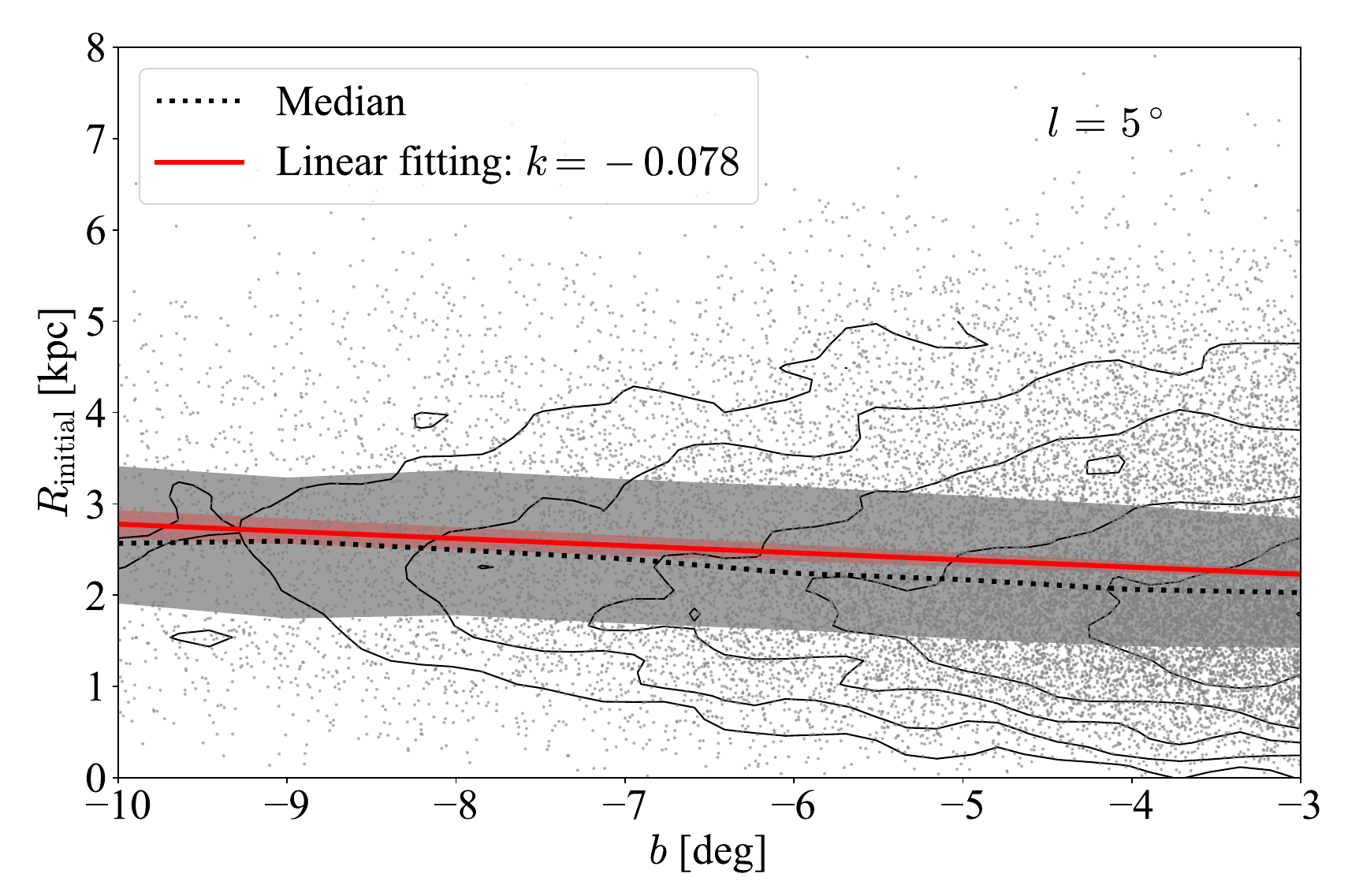}
\includegraphics[width=.495\textwidth, height=.28\textheight]{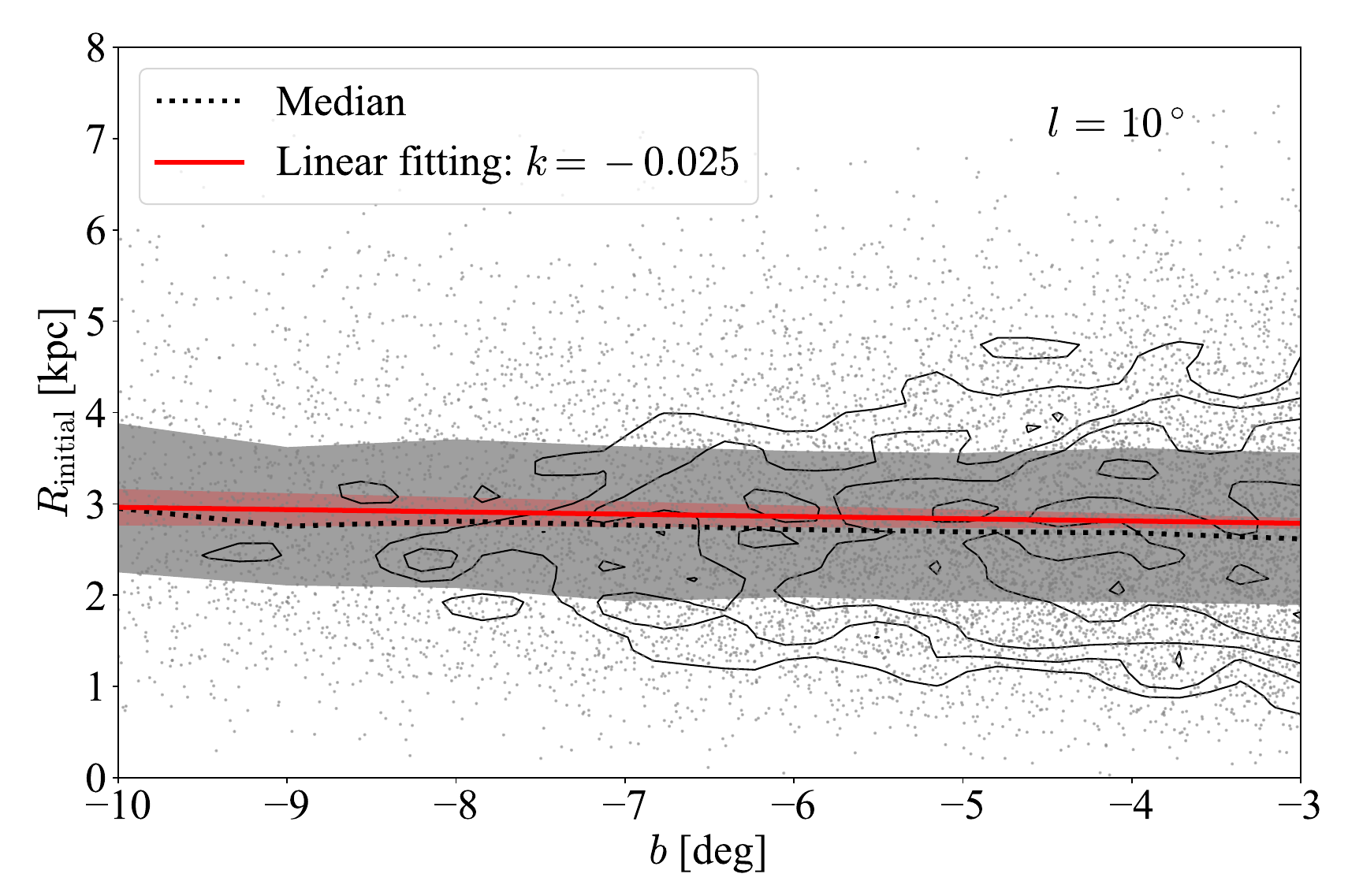}
\includegraphics[width=.495\textwidth, height=.25\textheight]{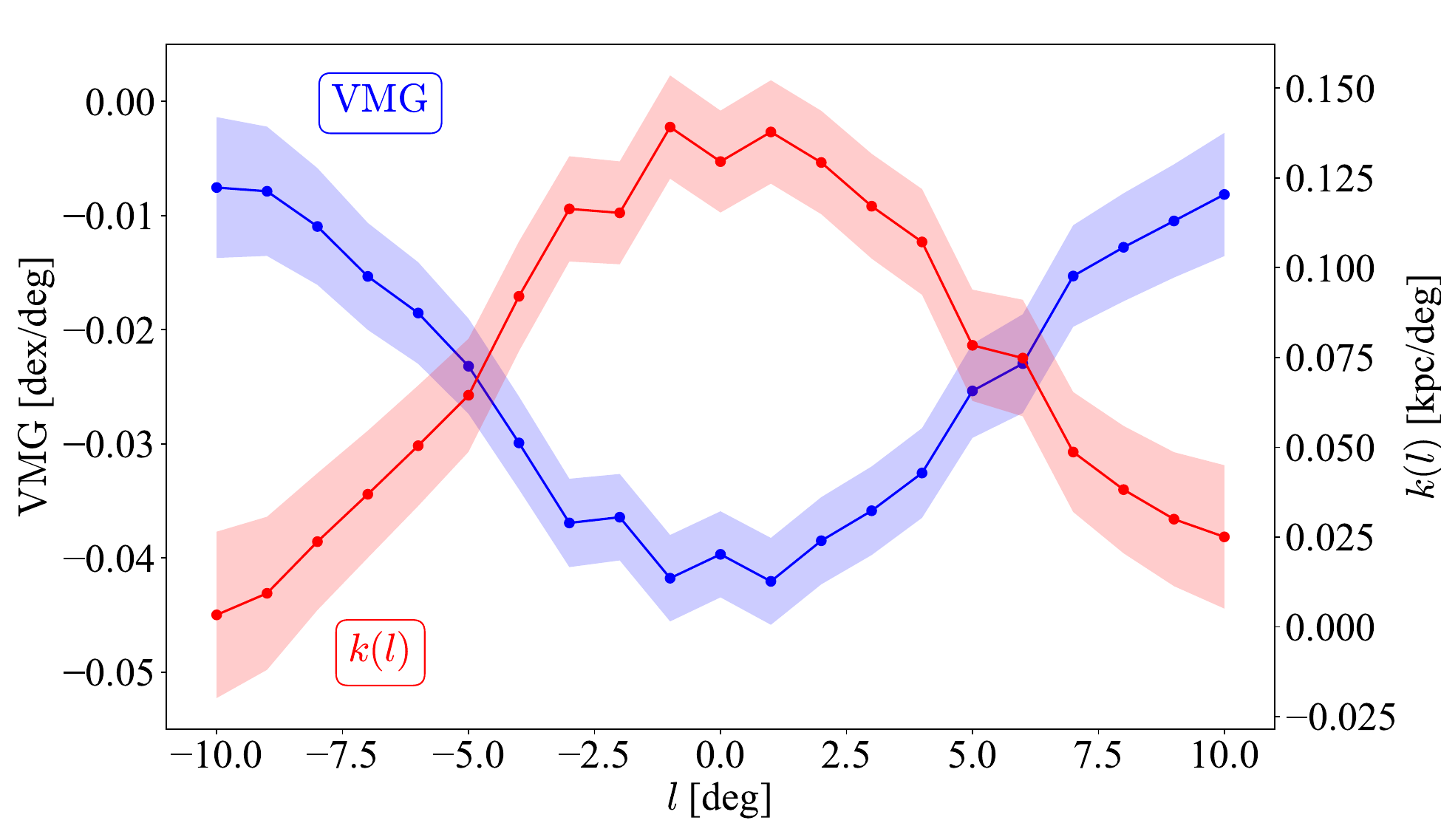}
\caption{The same as Figure \ref{fig: single ri vs. bf} but for the triple-disk model, and it is the superposition of all three disks shown.}
\label{fig: triple ri vs. bf}
\end{figure*}

\begin{figure*}[htbp!]
\centering
\includegraphics[width=0.325\textwidth, height=0.2\textheight]{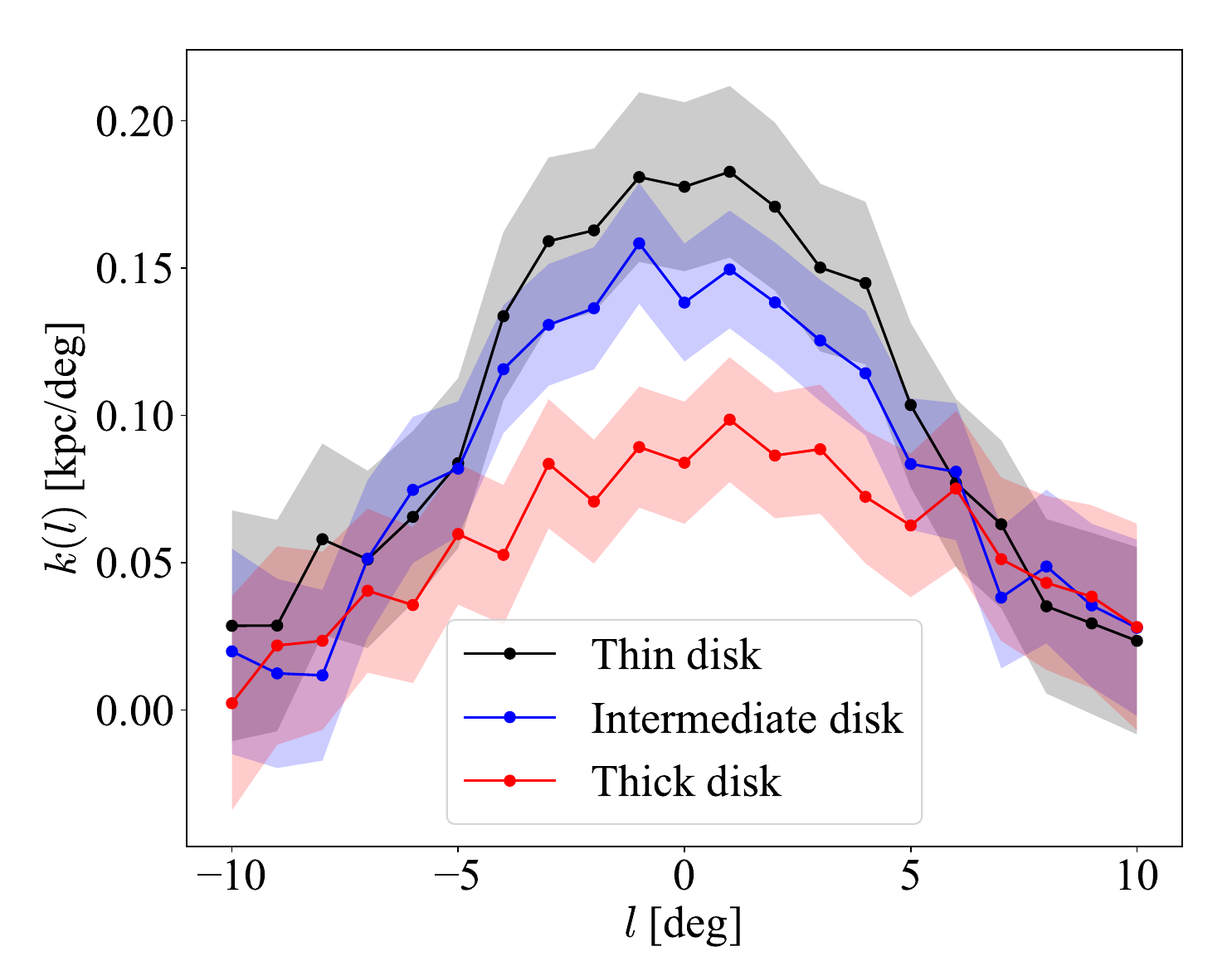}
\includegraphics[width=0.325\textwidth, height=0.2\textheight]{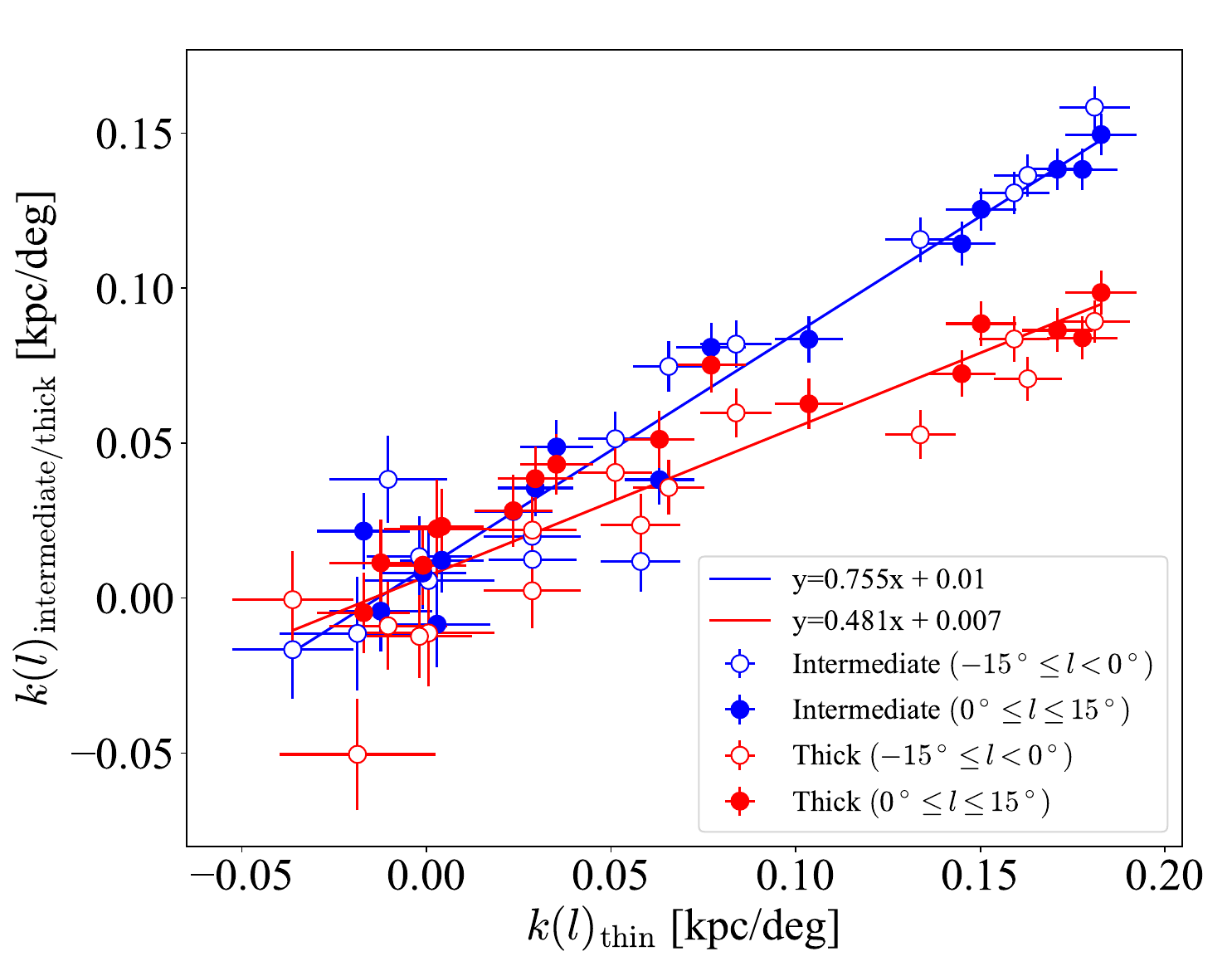}
\includegraphics[width=0.325\textwidth, height=0.2\textheight]{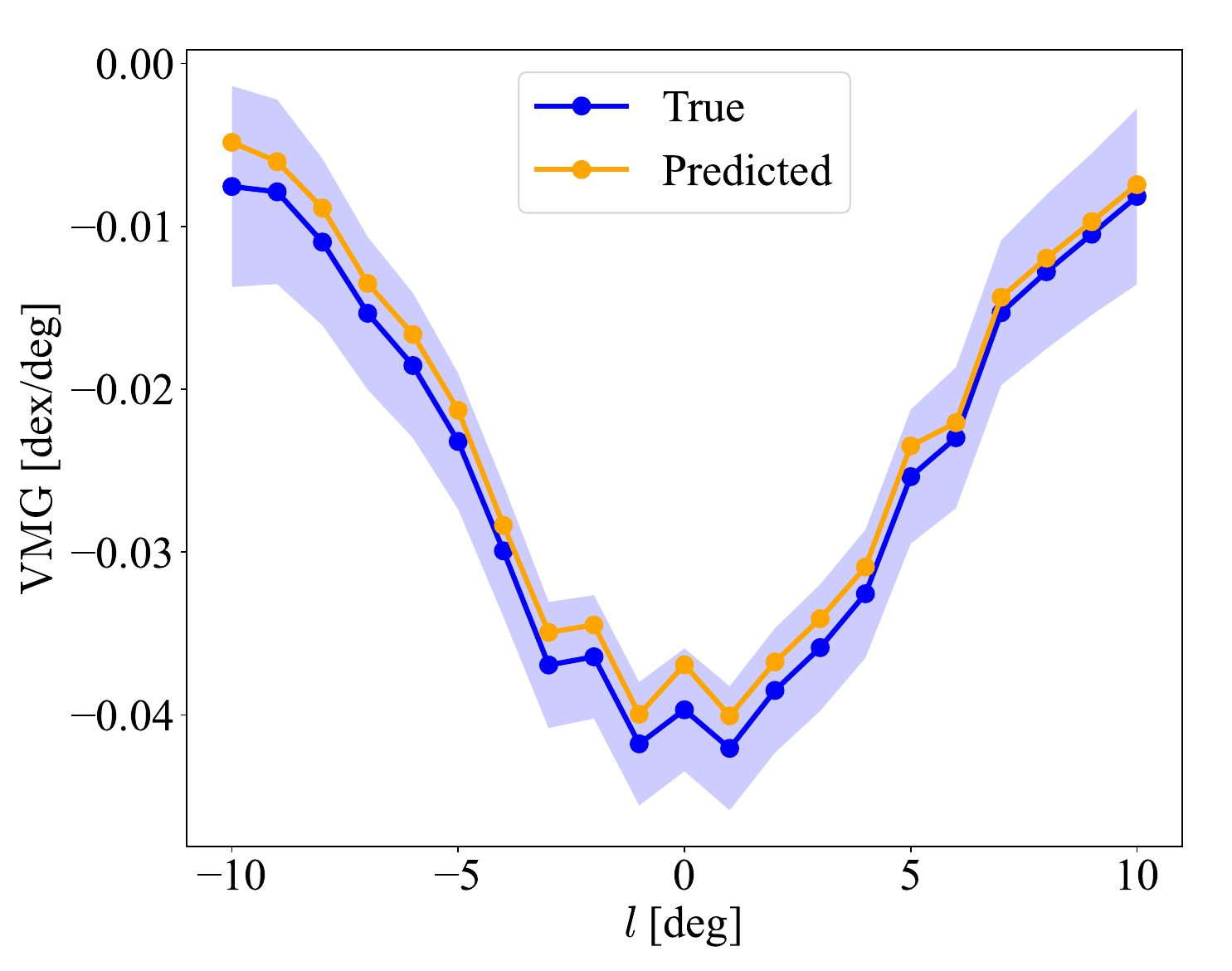}
\caption{Left panel: The $k(l)$ factors for each disk in the triple-disk model, similar to the red curves in the bottom right panels of Figure \ref{fig: single ri vs. bf} and Figure \ref{fig: triple ri vs. bf}. Middle panel: The $k(l)$ factors of the intermediate (blue) and thick (red) disks versus the ones of the thin disk, with the error bars showing the standard errors of the $k(l)$ factors. The blue and red straight lines are the linear fittings of the data points in the corresponding color, and their Pearson correlation coefficients are all greater than 0.9. Right panel: The VMG in the triple-disk model for longitudinal $1^\circ$ width stripes within $-10^\circ\leq l\leq 10^\circ$, for both the true values (blue, the same as the one in the bottom right panel of Figure \ref{fig: triple ri vs. bf}) and the predicted values from Equation \ref{eq: triple-disk formula} (yellow). The predicted values are consistent with the true values, with the deviations no more than $0.003\ \mathrm{dex}$. Note that the calculations of VMGs/$k(l)$ factors only use the bulge particles within $-10^\circ\leq b\leq -3^\circ$.}
\label{fig: triple heating factors}
\end{figure*}

\subsection{``Two-step heating" mechanism in the triple-disk model}\label{sec: triple: mechanism}

We observe that the ``two-step heating" mechanism (see \S~\ref{sec: single: mechanism}) also operates in the triple-disk model. Figure \ref{fig: triple two-step} shows the velocity dispersion profiles in the triple-disk model for both the entire model and each disk component at four epochs: the initial state, a pre-buckling bar epoch, an epoch near the saturation of buckling, and an epoch in a quasi-steady secular evolution state. The evolutionary track resembles the single-disk model: the bar instability induces the radial heating of particles (between the first and second columns in each panel), and the subsequent buckling instability results in the vertical heating of particles (between the second and third columns in each panel). 

Figure \ref{fig: triple ri vs. bf} shows that the redistribution of particles in the triple-disk model resembles the one in the single-disk model. Similar to Figure \ref{fig: single ri vs. bf} for the single-disk model, Figure \ref{fig: triple ri vs. bf} shows that the ``two-step heating" mechanism in the triple-disk model also redistributes particles in a nearly linear fashion. For the entire triple-disk model, the formed VMG is also linearly proportional to the $k(l)$ factors, indicated by their reflection symmetry in the bottom right panel of Figure \ref{fig: triple ri vs. bf}.

In particular, Figure \ref{fig: triple two-step} reveals that different disks undergo similar mixing and heating processes but with different strengths, reflecting the kinematic fractionation \citep{debatt_etal_2017} among the triple-disk model. To delve deeper, we present the distribution of $k(l)$ factors among the three disks in the left panel of Figure \ref{fig: triple heating factors}. As the $k(l)$ factors indicate, the mixing and heating processes are most pronounced in the thin disk and weakest in the thick disk, consistent with the kinematic fractionation scenario. Furthermore, the middle panel of Figure \ref{fig: triple heating factors} shows that the $k(l)$ factors in different disks are approximately linearly correlated. It indicates that the mixing and heating processes of the three disks are coupled with each other, as the onset of bar and buckling instabilities in the thin disk promotes the instabilities in the others. 

In summary, the ``two-step heating" mechanism also operates in a triple-disk model, and the heating processes among different disks are linearly correlated.

\subsection{Relationship between the VMG and the initial RMGs}\label{sec: triple: formula}

In the single-disk model, the initial RMG is linearly transformed into the VMG by the ``two-step heating" mechanism (see \S~\ref{sec: single: formula}). We observe that the three disks in the triple-disk model resemble this behavior. To understand such a linear transformation in more general cases, we generalize the single-disk model's Equation \ref{eq: single-disk formula} as follows:
\begin{equation} \label{eq: triple-disk formula}
\begin{split}
&\mathrm{VMG}(l)\equiv \dfrac{\partial \left<\mathrm{[Fe/H]}\right>}{\partial |b_\mathrm{final}|}\bigg|_l\\
&= \sum\limits_{\text{component }i} w_i \dfrac{\partial \left<\mathrm{[Fe/H]}\right>_i }{\partial |b_\mathrm{final}|}\bigg|_l\\
&= \sum\limits_{\text{component }i} w_i \dfrac{\partial \left<\mathrm{[Fe/H]}_0 + \alpha_R\times R_\mathrm{initial}\right>_i}{\partial |b_\mathrm{final}|}\bigg|_l\\
&= \sum\limits_{\text{component }i} w_i \left[\alpha_R\cdot\,k(l)\right]_i\ ,
\end{split}
\end{equation}
where $w_i$'s are the statistical weights for the disk components.

The right panel of Figure \ref{fig: triple heating factors} shows that Equation \ref{eq: triple-disk formula} can precisely predict the VMG in the triple-disk model. The true VMGs in the triple-disk model align well with the VMGs predicted by Equation \ref{eq: triple-disk formula}, where the statistical weights for the three disks are defined as their number fractions of bulge particles within $-10^\circ\leq b \leq -3^\circ$ at different longitudinal $1^\circ$ width strips. There are only slight differences between the true VMGs and the predicted VMGs. Therefore, for real galaxies, if there are multiple components with initial RMGs in the precursor disk, the ``two-step heating" mechanism will transform the RMGs into a VMG in the form of Equation \ref{eq: triple-disk formula}.

\section{Discussion}\label{sec: dis}

In the preceding sections, we construct single-disk and triple-disk models with initial RMGs. The two models successfully generate the VMG phenomenon, as reported in \cite{mar_ger_2013}. Such a VMG emerges naturally through a ``two-step heating" mechanism during the sequential bar and buckling instabilities (\S~\ref{sec: single: mechanism} and \S~\ref{sec: triple: mechanism}). In this section, we discuss the implications and limitations of these models.

\subsection{The ``two-step heating" mechanism is inherent in the formation of a BPX bulge}\label{sec: dis: inherent}

The ``two-step heating" mechanism found in \S~\ref{sec: single: mechanism} and \S~\ref{sec: triple: mechanism} are related to the sequential bar and buckling instabilities, which are inherent in the conventional formation process of a BPX bulge. In Appendix \ref{app: rigid halo}, we re-confirm the ``two-step heating" mechanism in a rigid halo galactic model, proving that the ``two-step heating" mechanism is independent of the dynamical friction between the bar and the DM halo. Therefore, the mechanism is independent of either the number of disk components or the dynamical friction between the bar and DM halo. The ``two-step heating" process is inevitable in any BPX bulge galaxy if its vertical thickening stems from the buckling instability as generally believed \citep{sel_ger_2020}.

If the precursor disk has an initial RMG (or RMGs), the ``two-step heating" mechanism will linearly transform the initial RMG(s) into a VMG (\S~\ref{sec: single: formula} and \S~\ref{sec: triple: formula}). Such a RMG is naturally expected in the inside-out formation scenario of galactic disks, which is now well-established \citep{mat_fra_1989, sharma_etal_2021, lu_etal_2022}. Therefore, it is reasonable to assume the Galactic precursor disk has a RMG or RMGs, and then the ``two-step heating" mechanism is a non-negligible aspect in forming the Galactic VMG.

\begin{table}
\caption{Parameters of the Gaussian metallicity\footnote{Note that these mean and dispersion values are slightly different from the model M1 of \cite{fragko_etal_2017}.}}\label{tab: Gaussian triple-disk}
\centering
\begin{tabular}{ccc}
\hline
Component & $\mu(\mathrm{[Fe/H]})$ & $\sigma(\mathrm{[Fe/H]})$
\\
\hline
Thin disk & $\tThinMHmean$ & 0.2 \\
Intermediate disk & $\tInterMHmean$ & 0.2 \\ 
Thick disk & $\tThickMHmean$ & 0.2 \\
\hline\noalign{\smallskip}
\end{tabular}
\end{table}

\subsection{VMG in the triple-disk model when tagged with radially independent Gaussian metallicity}\label{sec: disc: 3DG}

We admit that the VMG stemming from linear initial RMG(s) cannot reproduce all chemical properties of the Galactic bulge. \cite{dimatt_etal_2015} and \cite{fragko_etal_2017} reported some inconsistencies between observations and BPX bulges formed in a cold single-disk model with initial RMG. Such a model has a pathological BPX strength among different stellar components, lacking a positive longitudinal metallicity gradient (LMG) at the bulge mid-plane, and has other related pathological properties, as the cases in our single-disk and triple-disk models. To resolve these issues, \cite{fragko_etal_2017} construct their model M1, a triple-disk model with radially independent Gaussian metallicity distribution. Such a model can simultaneously reproduce the VMG and a positive LMG. To compare directly with their model M1, here we also discuss the properties of our triple-disk model when each disk is tagged with a radially independent Gaussian metallicity distribution as in their model M1. The parameters of the Gaussians are listed in Table \ref{tab: Gaussian triple-disk}, where the thin, intermediate, and thick disks are relatively metal-rich, metal-intermediate, and metal-poor, respectively. In this and the next subsections, we abbreviate the triple-disk model with the initial RMGs as the ``3DR model" and the triple-disk model with radially independent Gaussian metallicity as the ``3DG model" for convenience.

\begin{figure*}[htbp!]
\centering
\includegraphics[width=1.\textwidth, height=0.46\textheight]{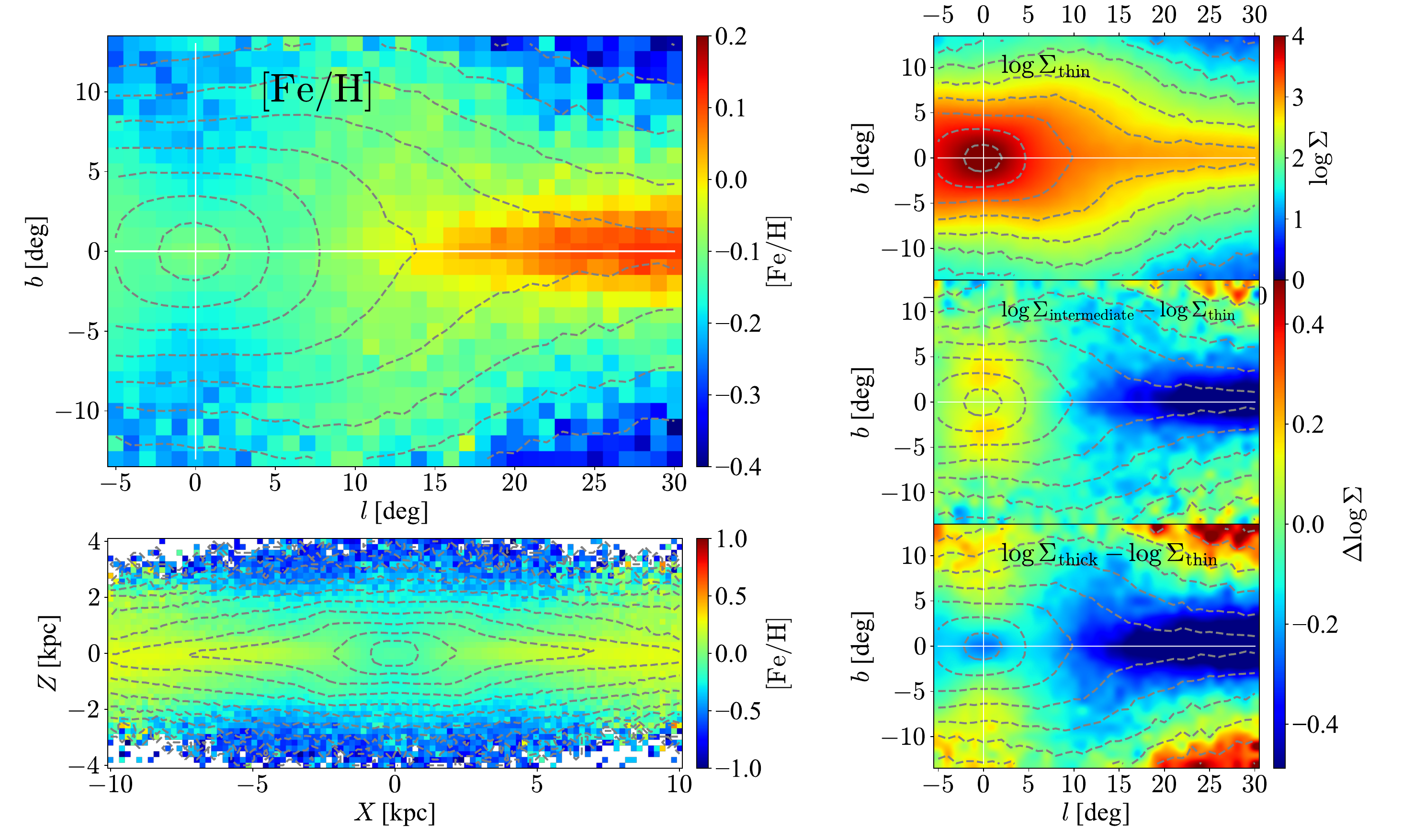}
\caption{Left column: The metallicity distribution of the 3DG model with grey contours indicating the stellar surface densities. The top panel shows the bulge metallicity distribution in $(l, b)$ coordinates, mirroring Figure 1 of \cite{fragko_etal_2017}. The bottom panel displays the side-on metallicity distribution, where the bar is parallel to the $X$-axis. This model exhibits properties similar to M1 of \cite{fragko_etal_2017} and Galactic observations \citep{wylie_etal_2021}, including a VMG throughout the bulge (about $-0.01\ \mathrm{dex/deg}$ at the bulge minor axis) and a positive longitudinal metallicity gradient in the mid-plane. The top panel in the right column shows the stellar surface density in the bulge of the thin disk. The middle and bottom panels in the right column show the excess surface densities of the intermediate and thick disks relative to the thin disk, and the grey contours in the background indicate the stellar surface density of the thin disk.}
\label{fig: triple Gauss tagged metal map}
\end{figure*}

In Figure \ref{fig: triple Gauss tagged metal map}, the left column shows the metallicity distribution of the 3DG model. In $(l,\ b)$ coordinates, the 3DG model produces a VMG similar to the model M1 of \cite{fragko_etal_2017}. Notably, it also exhibits a positive LMG in the mid-plane, seen in the top left panel. To understand how the VMG and positive LMG form in the 3DG model, we show the surface density of the thin disk and the excess surface densities of the intermediate and thick disks relative to the thin one in the right column of Figure \ref{fig: triple Gauss tagged metal map}. The thin disk's surface density dominates the mid-plane ($|b|\lesssim5^\circ$), especially in the outer region ($l\gtrsim10^\circ$), and the intermediate and thick disks dominate the off-plane region, especially in the inner part. Therefore, after the three disks are tagged with radially independent Gaussian metallicity, the negative VMG and positive LMG emerge naturally in the surface density-weighted metallicity distribution. Based on this explanation, galactic models with multiple components featuring suitable spatial variation and appropriate Gaussians of metallicity can yield similar metallicity patterns (i.e., the negative VMG and positive LMG).

However, some previous works have cautioned about the LMG. Such a LMG pattern is tightly related to the fact that the Galactic Bulge's metallicity at the near side is much higher than the GC and far side in observations, which may be ``an effect of the field viewing angles, which cause the nearer stars to be preferentially sampled closer to the plane than the farther stars" \citep{wylie_etal_2021}.

Besides, the 3DG model has two issues: (1) As a by-product of the positive LMG at the mid-plane, the metal-rich thin disk will always produce an inverted/positive RMG in the whole disk, as shown by the bottom left panel of Figure \ref{fig: triple Gauss tagged metal map}. Such an inverted RMG is inconsistent with the inside-out formation scenario of galactic disks and the Galactic observations \citep{bovy_etal_2019, haywoo_etal_2024}. An additional metal-rich component near the corotation resonance at the bar end \citep{chiapp_etal_2015} may be a possible reason for a positive LMG between $0^\circ \lesssim l \lesssim 30^\circ$ avoiding such an inverted RMG. (2) Its VMG at the bulge minor axis is only about $-0.01\ \mathrm{dex/deg}$, calculated by linearly fitting pixels that $|l|\leq0.5^\circ$ and $-10^\circ\leq b \leq-3^\circ$ in the top left panel of Figure \ref{fig: triple Gauss tagged metal map}. It is much shallower than the ones of the single-disk and 3DR models, and the observed value of the MW (approximately $-0.04\ \mathrm{dex/deg}$ in \citealt{gonzal_etal_2013}). The model M1 of \cite{fragko_etal_2017} also has a VMG shallower than the MW. Such a shallow gradient may be due to the absence of the VMG stemming from the initial RMGs under the ``two-step heating" mechanism. Therefore, the VMG stemming from the RMG(s) may still be important even for a multiple-component Galactic model. 

\begin{figure*}[htbp!]
\centering
\includegraphics[width=.83\textwidth, height=.3\textheight]{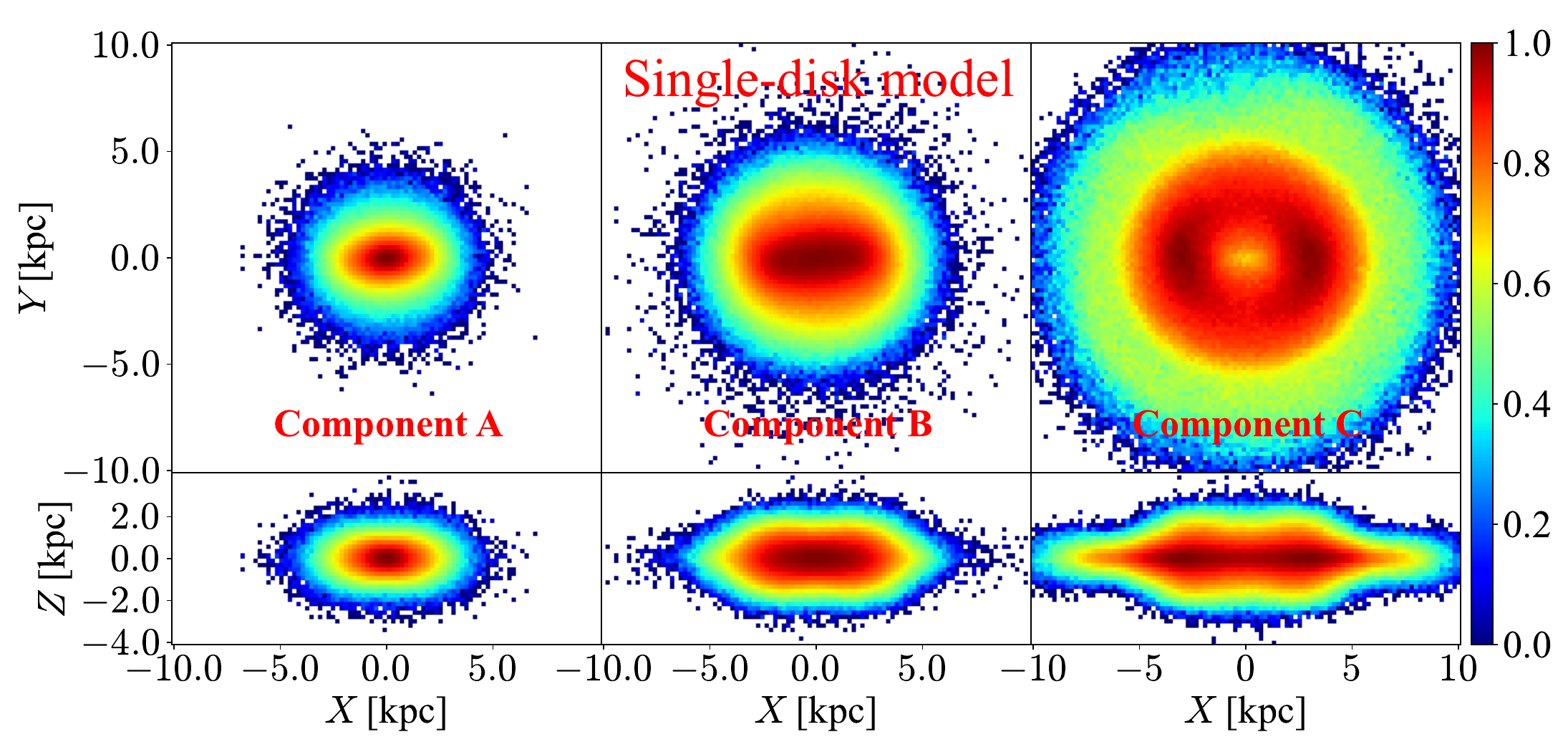}
\includegraphics[width=.83\textwidth, height=.3\textheight]{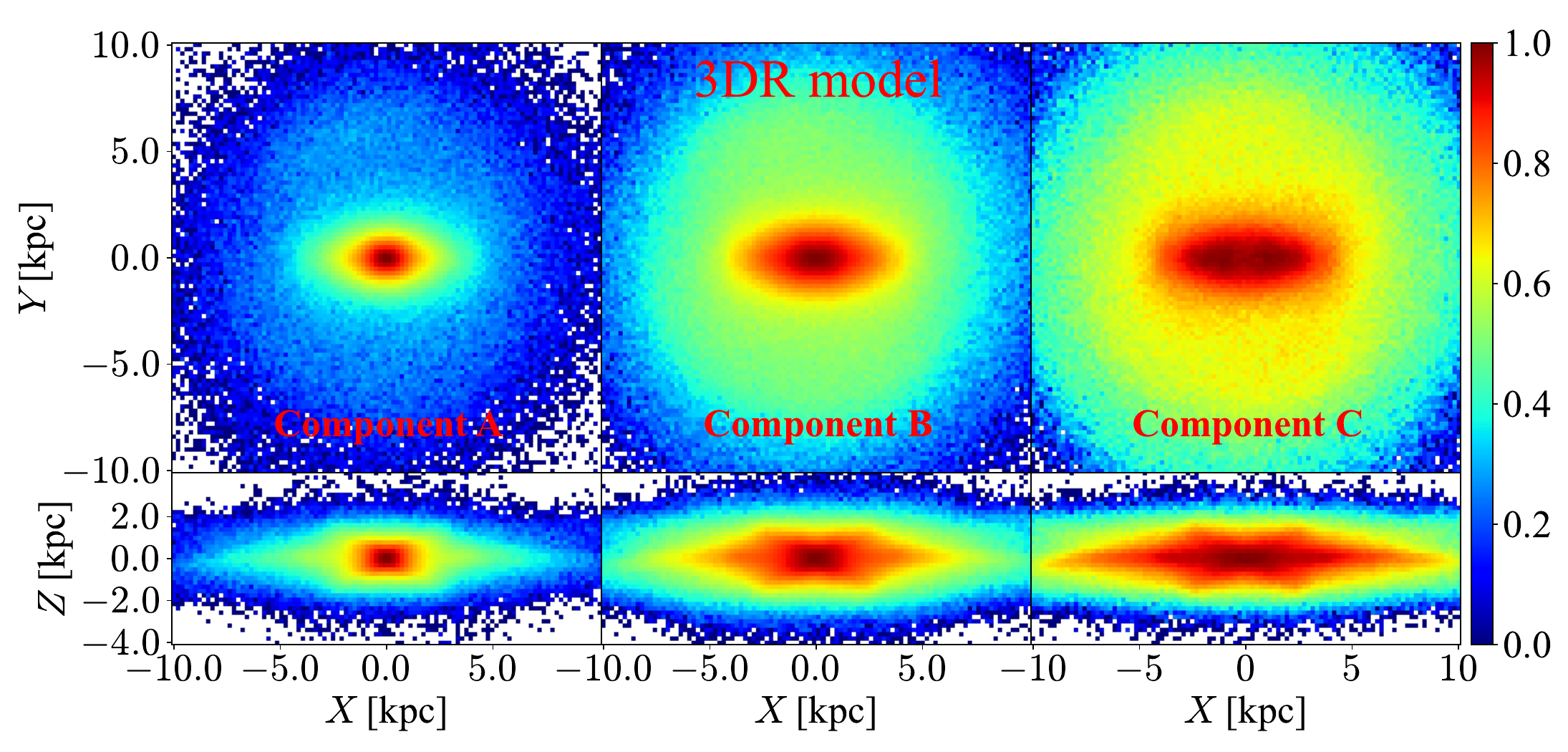}
\includegraphics[width=.83\textwidth, height=.3\textheight]{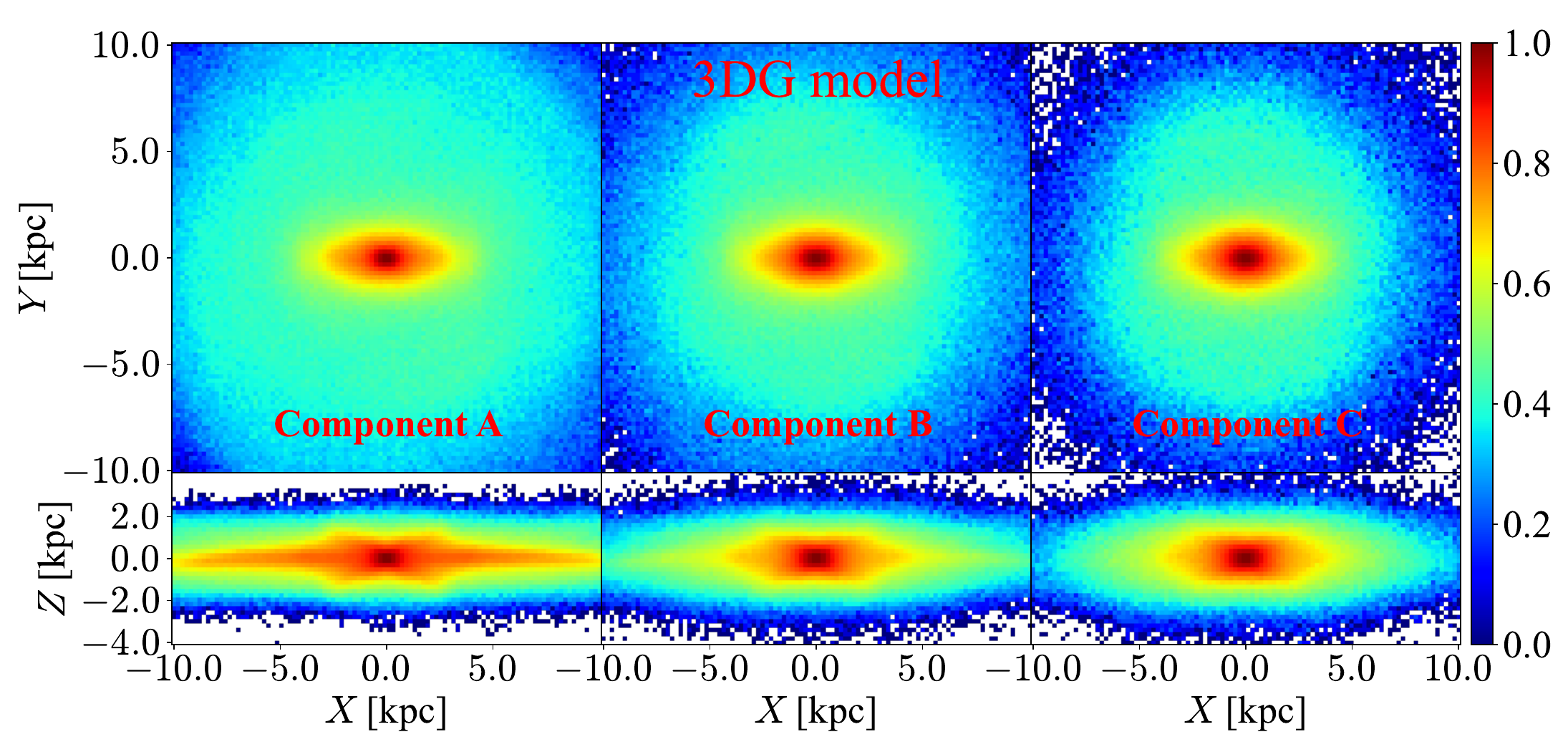}
\caption{The face-on and side-on views of the normalized logarithmic surface densities for the major Galactic components defined by \cite{ness_etal_2013_a} for the three models. From top to bottom: the single-disk model, the 3DR model, and the 3DG model. From left to right: Component A ($\mathrm{[Fe/H]}>0.0$), Component B ($-0.5<\mathrm{[Fe/H]}\leq0.0$), and Component C ($-1.0<\mathrm{[Fe/H]}\leq-0.5$). In the single-disk and 3DR models, the bar structure (face-on) and the BPX structure (side-on) are more pronounced in the more metal-poor components. In contrast, these structures are more pronounced in the more metal-rich components in the 3DG model.}
\label{fig: BPX bar}
\end{figure*}

\subsection{BPX strengths in different populations}\label{sec: dis: xshape}

Another critical issue of a BPX bulge forming in a cold single-disk model, as reported by \cite{dimatt_etal_2015}, is that the BPX structure is more pronounced in more metal-poor components. Such a trend is opposite to both the observations (e.g., \citealt{babusi_etal_2010, ness_etal_2012, uttent_etal_2012, rojas_etal_2014, catchp_etal_2016}) and cosmological zoom-in simulations in the literature, such as Auriga \citep{fragko_etal_2020, che_li_2022}. This discrepancy also exists in our single-disk model and 3DR model.

Figure \ref{fig: BPX bar} shows face-on and side-on views of the single-disk, 3DR and 3DG models for major components in the Galactic bulge defined by \cite{ness_etal_2013_a}: (1) Component A: $\mathrm{[Fe/H]}>0.0$, (2) Component B: $-0.5<\mathrm{[Fe/H]}\leq 0.0$, (3) Component C: $-1.0<\mathrm{[Fe/H]}\leq -0.5$. In the side-on views, the BPX structure is more pronounced in more metal-poor components for the single-disk and 3DR models, contradictory to observations \citep{dimatt_etal_2015}. The 3DG model shows an opposite trend which is more consistent with the observations.

Such an inconsistency stems from the different origins of the three components in these models. The three components in the single-disk and 3DR models have different initial radii due to the initial RMGs in the precursor disks. The more metal-poor components are initially outer and colder in the precursor disk. Therefore, according to the ``two-step heating" mechanism (\S~\ref{sec: single: mechanism} and \S~\ref{sec: triple: mechanism}), they undergo more violent heating and mixing, resulting in a more pronounced BPX shape. However, in the 3DG model, the three components, dominated by the thin, intermediate, and thick disks, respectively (see Table \ref{tab: Gaussian triple-disk}), are all distributed across the whole disk and then undergo similar dynamical evolution. Thus, according to the kinematic fractionation scenario \citep{debatt_etal_2020}, the BPX bulge in the 3DG model is more pronounced in more metal-rich components, as they are dynamically colder at the beginning.

The pathological trends in the single-disk and 3DR models are partially because the overly simplified radially tagged metallicity overestimates the initial RMGs. When trying to match the metallicity distribution of the single-disk and 3DR models to the observations \citep{gonzal_etal_2013}, we have to attribute all Galactic metallicity gradients to the initial RMGs, as in Equation \ref{eq: single-disk formula} and \ref{eq: triple-disk formula}. Consequently, to get a VMG compatible with the MW we must assign some large absolute values to the initial RMGs in these models (see Table \ref{tab: single-disk} and \ref{tab: ri based triple-disk}). As a result, these models have very metal-rich cores at their galactic center (see Figure \ref{fig: single metal map} and \ref{fig: triple ri tag metal map}), and their metallicity decreases dramatically away from such cores, resulting in very metal-poor outer skirts.

The lack of chemical enrichment during the $N$-body simulation and the intrinsic multiple components in the precursor disks at different radii may be the physical reasons why the initial RMGs are unavoidably overestimated. The chemical enrichment is missing in the radially tagged metallicity but somewhat mimicked by the multiple components in the 3DG model, whose averages metallicity are located in the Galactic major components' metallicity range \citep{ness_etal_2013_a}, accounting for why the BPX strength in each component is compatible with observations. Besides the 3DG model's approach, \cite{debatt_etal_2020} found that the initial actions $(J_R,\ J_\phi,\ J_Z)$-based metallicity tagging, rather than a simple linear RMG based solely on $R_\mathrm{initial}$ as in Equation \ref{eq: ri based tagging}, can also resolve the issue of BPX strength. This is because the $J_R$ and $J_Z$ define sub-populations inside the $J_\phi$-defined stellar populations at different radii. \cite{debatt_etal_2020} also found that solely $J_{\phi}$-based metallicity, similar to our $R_\mathrm{initial}$-based metallicity, still suffers from the BPX strength issue, which indicates the necessity of multiple components at all radii in the precursor disk to produce a trend of BPX strength compatible with observations.

\subsection{Summary}\label{sec: dis: summary}

The ``two-step heating" mechanism is associated with two inherent dynamical instabilities during the conventional formation process of a BPX bulge. It linearly transforms the initial RMGs in the precursor disks into a VMG (see \S~\ref{sec: single: formula} and \S~\ref{sec: triple: formula}). If all Galactic metallicity gradients are attributed to it, the initial RMG(s) will be unavoidably over-estimated (\S~\ref{sec: dis: xshape}). More disk components or a more delicate metallicity tagging may improve the match between the models and observations. However, although the dynamical mechanism is imperfect, ignoring its effect on shaping VMG will produce a VMG shallower than the MW (\S~\ref{sec: disc: 3DG}). A complete explanation of the Galactic metallicity gradients should also synthetically explore other tagging possibilities, the multiple-component nature of the Galaxy, and chemical enrichment during the Galactic evolutionary history. These considerations are beyond the scope of this paper, so we leave them for future study.

\section{Conclusions}\label{sec: con}

In this study, we construct a single-disk and a triple-disk $N$-body models to explore the dynamical origin of the Galactic vertical metallicity gradient (VMG). The disks in the two models, similar to \cite{mar_ger_2013}, have initial radial metallicity gradients. They successfully generate the observed VMG pattern. Through kinematic analyses, we demonstrate that the VMG in these models forms through a ``two-step heating" mechanism driven by the sequential bar and buckling instabilities (\S~\ref{sec: single: mechanism} and \S~\ref{sec: triple: mechanism}), after which more metal-poor particles end up at greater vertical heights. This mechanism linearly transforms the initial radial metallicity gradient into the VMG (\S~\ref{sec: single: formula} and \S~\ref{sec: triple: formula}) where the linear coefficient $k(l)$ encodes the heating strength during the evolutionary history. Especially in the triple-disk model, the three disks' $k(l)$ factors are approximately linearly correlated. Through comparison with previous works, we also discuss the limitations of the ``two-step heating" mechanism on shaping the VMG in Section \ref{sec: dis}. We find that although the ``two-step heating" mechanism is imperfect, it is still a non-negligible factor in the formation of the Galactic VMG.

In summary, we find that:
\begin{enumerate}[itemsep=0pt, topsep=0pt, parsep=0pt]
\item A ``two-step heating" mechanism during the secular evolution of a BPX bulge (\S~\ref{sec: single: mechanism} and \S~\ref{sec: triple: mechanism}).
\item If an initial radial metallicity gradient exists, it will be transformed into a VMG by the ``two-step heating" mechanism in a linear fashion (Equation \ref{eq: single-disk formula} and \ref{eq: triple-disk formula}).
\item In a multiple-disk model, the ``two-step heating" mechanism in each disk is coupled (\S~\ref{sec: triple: mechanism}).
\end{enumerate}

A complete explanation of the Galactic metallicity distribution in future research should also consider chemical enrichment and the impact of wet mergers during the Galactic evolution, which we leave for future works.

\begin{acknowledgments}
We sincerely thank Ortwin Gerhard, Paola Di Matteo, Zhi Li, Zhao-Yu Li, and Sandeep K. Kataria for their helpful discussions. We also thank the referee for helpful comments and suggestions. The research presented here is partially supported by the National Center for High-Level Talent Training in Mathematics, Physics, Chemistry, and Biology; by the National Key R\&D Program of China under grant No. 2018YFA0404501; by the National Natural Science Foundation of China under grant No. 12025302, 11773052, 12103031; by the Initiative Postdocs Supporting Program (No. BX2021183); by the ``111'' Center of the Ministry of Education of China under grant No. B20019; and by the China Manned Space Project under grant No. CMS-CSST-2021-B03. This work made use of the Gravity Supercomputer at the Department of Astronomy, Shanghai Jiao Tong University, and the facilities of the Center for High Performance Computing at Shanghai Astronomical Observatory.
\end{acknowledgments}

\appendix
\section{``Two-step heating" Mechanism in a Rigid Halo Model}\label{app: rigid halo}

\begin{figure*}[htbp!]
\centering
\includegraphics[width=1.\textwidth, height=.45\textheight]{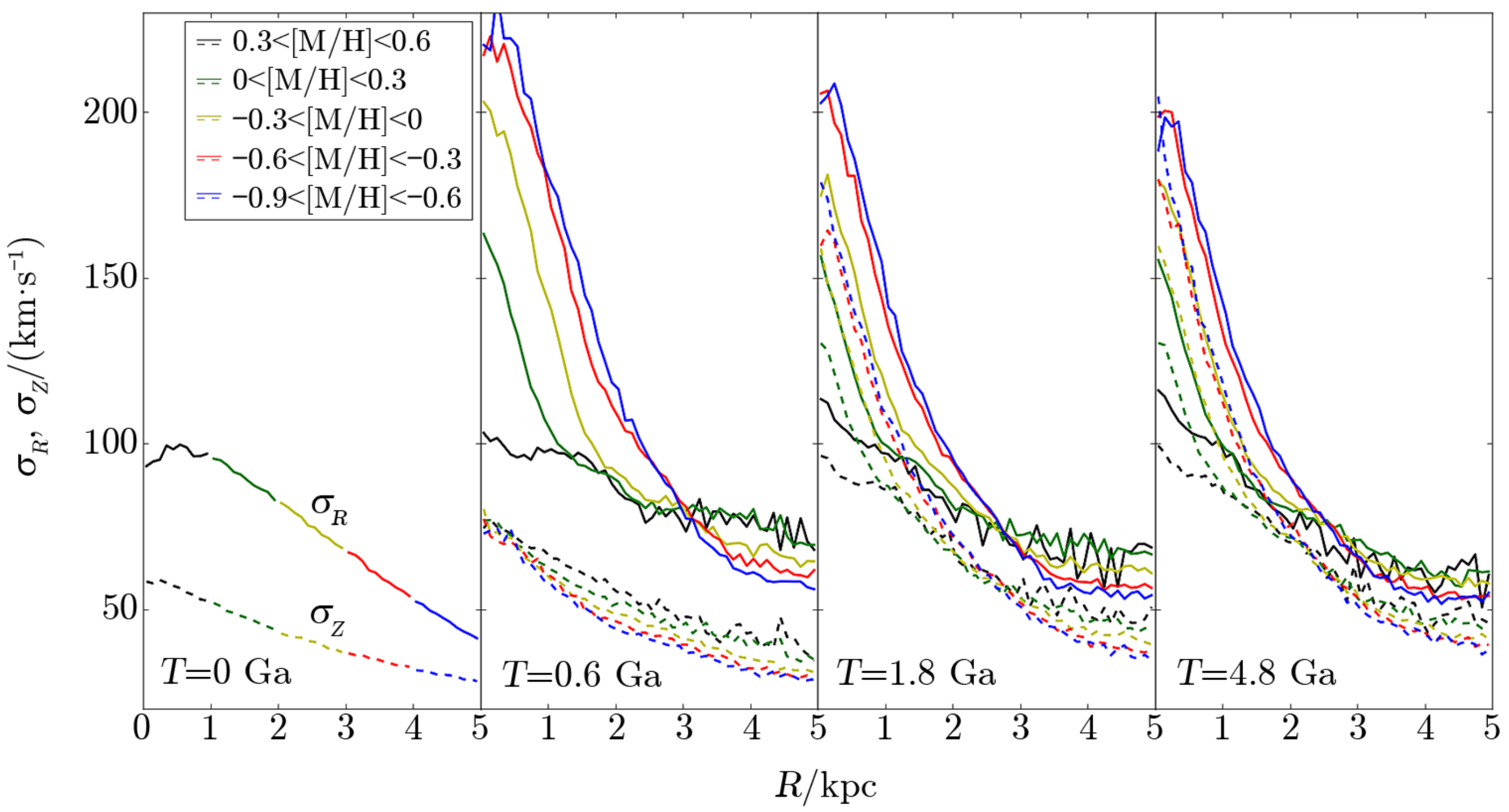}
\caption{The cylindrical radial (solid lines) and vertical (dashed lines) velocity dispersion profiles of a rigid halo model taken from \cite{shen_etal_2010}. The epoch of each panel from left to right: the initial snapshot at 0 $\mathrm{Gyr}$, bar formed but before buckling at 0.6 $\mathrm{Gyr}$, after the buckling at 1.8 $\mathrm{Gyr}$, and a snapshot in quasi-steady state at 4.8 $\mathrm{Gyr}$. Particles in different 1 $\mathrm{kpc}$ initial radii bins are color-coded with different colors, from inside to outside: black, green, yellow, red, and blue. Similar to the live halo single-disk model in the main context, the dispersion profiles of the rigid halo model also show significant variation from the first column to the second and from the second to the third. These variations indicate the violent radial and vertical heating processes during the bar and buckling instabilities. This figure is extracted from \cite{liu_she_2023} with permission and corresponds to our Figure \ref{fig: single two-step} in the main context.}
\label{fig: two-step in shen10}
\end{figure*}

Figure \ref{fig: two-step in shen10} shows the evolutionary track of velocity dispersions in the model of \cite{shen_etal_2010}, similar to Figure \ref{fig: single two-step}. The first three panels also correspond to the characteristic epochs separated by the bar and buckling instabilities, and the last panel is in an epoch of quasi-steady secular evolution. The overall pattern matches Figure \ref{fig: single two-step}, suggesting that the ``two-step heating" mechanism is also effective in a rigid halo galactic model.

\bibliographystyle{aasjournal}
\bibliography{references}
\end{document}